\documentclass[twocolumn]{aastex701}

\submitjournal{The Astrophysical Journal}
\accepted{February 5, 2026}

\shorttitle{RR Lyrae in WLM}
\shortauthors{Slaughter et al.}

\usepackage{amsmath}

\begin{document}

\title{The JWST Resolved Stellar Populations Early Release Science Program. IX. The RR Lyrae Population in WLM with HST and JWST}


\author[0000-0002-5752-3780]{Catherine M.\ Slaughter}
\email{slaug098@umn.edu}
\affiliation{Minnesota Institute for Astrophysics, University of Minnesota, Minneapolis, MN 55455, USA}

\author[0000-0003-0605-8732]{Evan D.\ Skillman}
\email{skill001@umn.edu}
\affiliation{Minnesota Institute for Astrophysics, University of Minnesota, Minneapolis, MN 55455, USA}

\author[0000-0002-1445-4877]{Alessandro Savino}
\email{asavino@berkeley.edu}
\affiliation{Department of Astronomy, University of California, Berkeley, Berkeley, CA 94720, USA}

\author[0000-0002-6442-6030]{Daniel R.\ Weisz}
\email{dan.weisz@berkeley.edu}
\affiliation{Department of Astronomy, University of California, Berkeley, Berkeley, CA 94720, USA}

\author[0000-0001-7531-9815]{Meredith Durbin}
\email{meredith.durbin@berkeley.edu}
\affiliation{Department of Astronomy, University of California, Berkeley, Berkeley, CA 94720, USA}

\author[0000-0003-2861-3995]{Jay Anderson}
\email{jayander@stsci.edu}
\affiliation{Space Telescope Science Institute, 3700 San Martin Drive, Baltimore, MD 21218, USA}

\author[0000-0003-4850-9589]{Martha L.\ Boyer}
\email{mboyer@stsci.edu}
\affiliation{Space Telescope Science Institute, 3700 San Martin Drive, Baltimore, MD 21218, USA}

\author[0000-0002-2970-7435]{Roger E.\ Cohen}
\email{rc1273@physics.rutgers.edu}
\affiliation{Department of Physics and Astronomy, Rutgers, the State University of New Jersey,  136 Frelinghuysen Road, Piscataway, NJ 08854, USA}

\author[0000-0003-0303-3855]{Andrew A.\ Cole}
\email{andrew.cole@utas.edu.au}
\affiliation{School of Natural Sciences, University of Tasmania, P.O. Box 807, Sandy Bay, Tasmania 7006, Australia}

\author[0000-0001-6464-3257]{Matteo Correnti}
\email{correnti@stsci.edu}
\affiliation{INAF Osservatorio Astronomico di Roma, Via Frascati 33, 00078, Monteporzio Catone, Rome, Italy}
\affiliation{ASI-Space Science Data Center, Via del Politecnico, I-00133, Rome, Italy}

\author[0000-0001-8416-4093]{Andrew E.\ Dolphin}
\email{adolphin@rtx.com}
\affiliation{Raytheon, 1151 E. Hermans Rd.,Tucson, AZ 85756}
\affiliation{Steward Observatory, University of Arizona, 933 N. Cherry Avenue, Tucson, AZ 85719, USA}

\author[0000-0002-7007-9725]{Marla C.\ Geha}
\email{marla.geha@yale.edu}
\affiliation{Department of Astronomy, Yale University, New Haven, CT 06520, USA}

\author[0000-0002-5581-2896]{Mario Gennaro}
\email{gennaro@stsci.edu}
\affiliation{Space Telescope Science Institute, 3700 San Martin Drive, Baltimore, MD 21218, USA}
\affiliation{The William H.\ Miller III Department of Physics \& Astronomy, Johns Hopkins University, 3400 N.\ Charles Street, Baltimore, MD 21218, USA}

\author[0000-0002-3204-1742]{Nitya Kallivayalil}
\email{njk3r@virginia.edu}
\affiliation{Department of Astronomy, University of Virginia, 530 McCormick Road, Charlottesville, VA 22904, USA}

\author[0000-0001-6196-5162]{Evan N.\ Kirby}
\email{ekirby@nd.edu}
\affiliation{Department of Physics and Astronomy, University of Notre Dame,
225 Nieuwland Science Hall, Notre Dame, IN 46556, USA}

\author[0000-0001-5538-2614]{Kristen B.\ W.\ McQuinn}
\email{kmcquinn@stsci.edu}
\affiliation{Department of Physics and Astronomy, Rutgers, the State University of New Jersey,  136 Frelinghuysen Road, Piscataway, NJ 08854, USA}
\affiliation{Space Telescope Science Institute, 3700 San Martin Drive, Baltimore, MD 21218, USA}

\author[0000-0002-8092-2077]{Max J.\ B.\ Newman}
\email{mjbnewman25astro@gmail.com}
\affiliation{Department of Physics and Astronomy, Rutgers, the State University of New Jersey,  136 Frelinghuysen Road, Piscataway, NJ 08854, USA}
\affiliation{Space Telescope Science Institute, 3700 San Martin Drive, Baltimore, MD 21218, USA}

\author[0000-0003-1634-4644]{Jack T.\ Warfield}
\email{jtw5zc@virginia.edu}
\affiliation{Department of Astronomy, University of Virginia, 530 McCormick Road, Charlottesville, VA 22904, USA}

\author[0000-0002-7502-0597]{Benjamin F.\ Williams}
\email{benw1@uw.edu}
\affiliation{Department of Astronomy, University of Washington, Box 351580, U.W., Seattle, WA 98195-1580, USA}


\begin{abstract}

RR Lyrae stars are a common, dependable Population II distance indicator, and provide an independent tracer of early star formation. Here, we utilize archival HST/ACS and JWST/NIRCam observations of the nearby dwarf star-forming galaxy WLM to study RR Lyrae in JWST filters. We independently identify RR Lyrae in HST and JWST imaging in order to evaluate JWST’s efficacy at characterizing RR Lyrae in the near-IR. We use an MCMC template-fitting technique to obtain periods, amplitudes, and mean magnitudes from the RR Lyrae time-series data. The spatially overlapping HST and JWST observations allow us to directly compare the same sources observed with the instruments, and calibrate the NIRCam F090W and F150W RR Lyrae period-Wesenheit-metallicity (PWZ) relation to the \textit{Gaia}-consistent HST PWZ. We additionally assess the epoch-to-epoch consistency of NIRCam photometry, and find evidence of burn-in. We conclude that the zero-point offset is negligible compared to the uncertainties from the template fitting. We conduct an MCMC fit of the PWZ with both HST and JWST data. Our results are three-fold. First, we find that we can reliably identify RR Lyrae in NIRCam data, but light-curve template fitting proves difficult on short-baseline observations. Second, the HST PWZ fit yields a distance modulus to WLM of $\mu = 24.85\pm0.05$ ($0.93\pm0.02$ Mpc). This is closer than previous measurements, primarily attributed to consistency with the \textit{Gaia} scale. Lastly, although the JWST PWZ fit has large uncertainties and a poorly-constrained slope, it represents a first-of-its-kind PWZ calibration in NIRCam filters.

\end{abstract}

\section{Introduction} \label{sec:intro}
\subsection{Variable Stars in Resolved Stellar Population Observations}\label{sec:intro_VS}

The Hubble Space Telescope (HST) has allowed us to measure the distances, lifetime star formation histories (SFHs), and age-metallicity relationships of a significant number of galaxies within the environs of the Local Group through the production of deep color-magnitude diagrams (CMDs)  \citep[e.g.,][]{Brown2003,Brown2006,Monelli2010a,Monelli2010b,Hidalgo2011,Skillman2017}.
Typically, because of the long integration times required, these studies have been accompanied by separate studies of the galaxy's short-period variable star populations, including, most importantly, studies of their RR Lyrae stars \citep[e.g.,][]{Brown2004,Bernard2009,Bernard2013,McQuinn2015,Martinez2017}.
The RR Lyrae stars provide independent measurements of distances and an independent constraint on the oldest star formation rate \citep[e.g.,][]{Bernard2008,Martinez2017,Savino2015,Savino2019} because the majority of RR Lyrae are expected to be at least 10 Gyr old \citep[][]{Walker1989, Savino2020, Bobrick2024}.
It has been proposed that combining observations of RR Lyrae stars with CMDs which do not reach down to the oldest main sequence turnoffs may provide lifetime SFHs at a greatly reduced cost in telescope time \citep[e.g.,][]{Monelli2022}. 

We expect that JWST observations of stellar populations in nearby galaxies will build upon the legacy from HST. Recently, \citet{Wang2025} conducted a study of several classes of periodic variable sources (including RR Lyrae stars) in JWST Large Magellanic Cloud (LMC) archival data with sparse and irregular cadence.
Here we analyze the RR Lyrae variable stars in the JWST early release science program \citep[ERS;][]{Weisz2023}  
NIRCam observations of the nearby dwarf star-forming galaxy WLM.  One of the two 
NIRCam fields was placed co-spatially with an existing deep HST ACS field.  This allows the direct comparison of HST and JWST observations of the same RR Lyrae stars, as well as a distance determination consistent with the \textit{Gaia} calibrated distance scale \citep[e.g.,][]{Savino2022}.  
Through this analysis, we will determine the best way to observe and analyze JWST 
observations of RR Lyrae stars in nearby galaxies and calibrate the JWST observations to the HST observations.

\begin{figure}
    \centering
    \includegraphics[width=\linewidth]{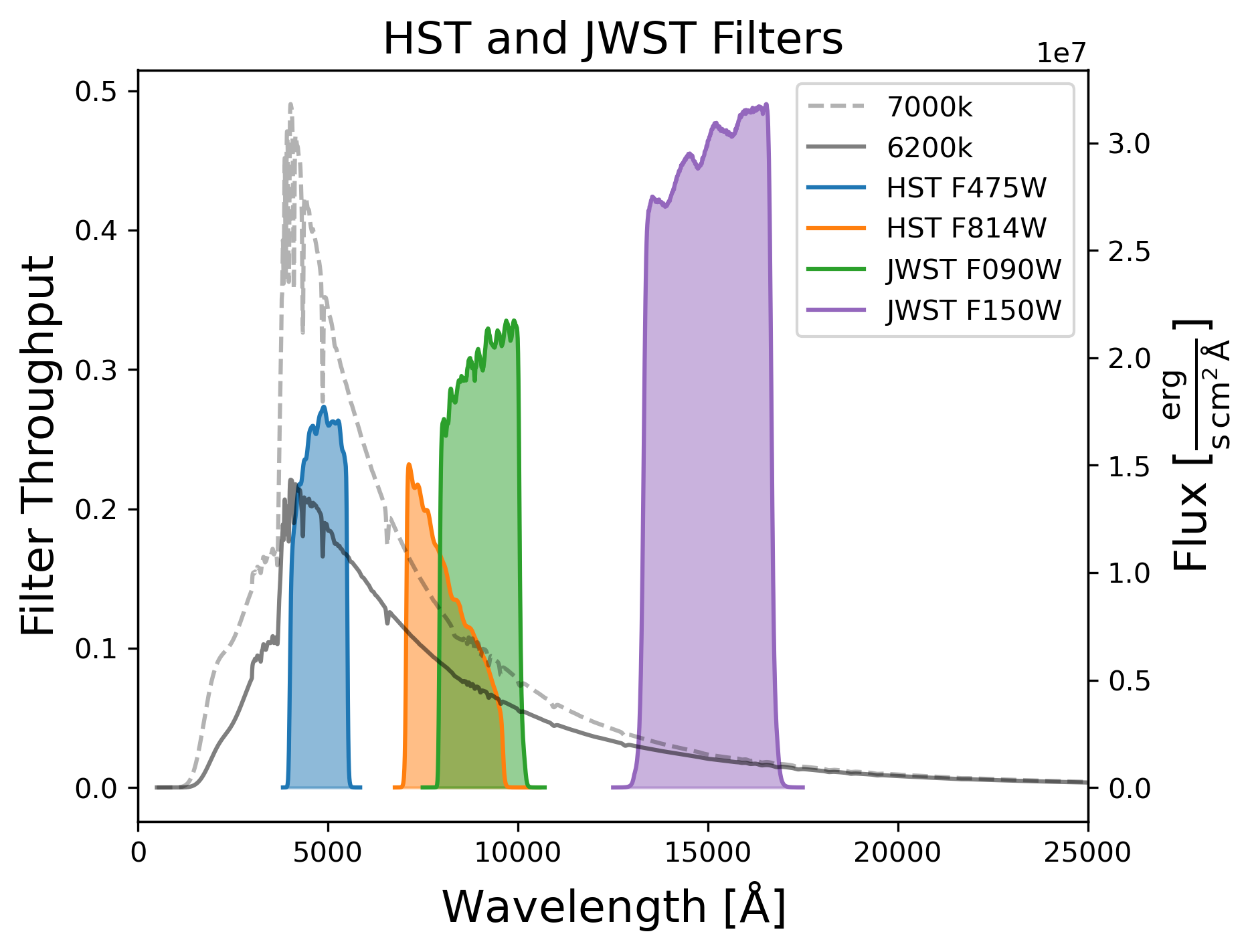}
    \caption{Transmission curves for the four filters used in this study, overlaid on synthetic \texttt{PHOENIX} spectra (smoothed for readability purposes) for two broadly representative maximum (dashed, light gray) and minimum (solid, dark gray) RR Lyrae temperatures. For illustrative purposes, the difference in flux between these spectra in a given wavelength range serves as a rough proxy for the expected amplitude of an RR Lyrae for that filter. As such, we can see how a change in effective temperature corresponds to a much larger change in flux in the bluest filter (HST F475W), compared to the other three used in this study. Similarly, we expect relatively small amplitudes in the reddest filter (JWST F150W). Note the overlap in wavelength coverage for the HST F814W and JWST F090W filters.}
    \label{fig:filters}
\end{figure}

There are several potential challenges to observing RR Lyrae stars in nearby galaxies with JWST archival data.
First, the amplitudes of the variations in RR Lyrae stars decrease with increasing wavelength \citep[e.g.,][Fig.\ 4]{Monson2017}. To illustrate this effect, we plot the pass-bands for the four filters used in this study, overlaid on example \texttt{PHOENIX} synthetic spectra \citep{Husser2013} for two broadly representative maximum and minimum RR Lyrae temperatures in \autoref{fig:filters}. 
Second, for JWST, in order to minimize the wear on the  moving parts in NIRCam's filter wheel, switching back and forth between filters while continuously observing a target is strongly discouraged. This was regularly done with HST in order to spread epochs in a single filter over the maximum time duration, optimizing phase coverage over a range in periods (see \autoref{sec:data}).
Third, the data available from the JWST standard pipeline do not produce images suitable for recovering variable sources (see \autoref{sec:TSO}).
And, finally, due to the superior sensitivity of JWST relative to HST, observations to study stellar populations will be shorter, providing less phase coverage for variable stars.
It should be noted that the final three concerns listed are of particular relevance when conducting research using \textit{archival} observations originally taken for resolved population studies, as we do here. In contrast, the first challenge stems from the physics of RR Lyrae pulsation and the JWST filter set, and would not be meaningfully alleviated by changing observing program specifications.

\subsection{The Nearby Dwarf Irregular Galaxy WLM}

WLM is a dwarf irregular galaxy at a distance of $\approx$ $0.98$ Mpc \citep{Lee2021}. The compilation of \citet{mcConnachie2012} provides a stellar mass of $4.3\times 10^7$ M$_{\odot}$ \citep{Kepley2007, Leaman2009} and a stellar mean metallicity $\langle\textrm{[Fe/H]}\rangle = -1.27$ \citep{Leaman2009}.  

The distance to WLM is well constrained, with studies in the literature demonstrating good agreement across several methods.  An extensive study of the distance to WLM has been conducted by \citet{Lee2021}; we provide a few historical highlights here.  A Tip of the Red Giant Branch (TRGB) distance was first measured in \citet{Lee1993} with a distance modulus of $\mu_{I} = 24.87 \pm 0.08$ mag. This was followed by TRGB measurements from \citet{McConnachie2005}, \citet{Rizzi2007}, and \citet{Pietrzynski2007} yielding 
25.84 $\pm$ 0.08 mag, 24.93 $\pm$ 0.04 mag, and 24.91 $\pm$ 0.08 mag.
\citet{Lee1993} also reported an I-band Cepheid distance modulus of 24.92 mag. 
Later, \citet{Gieren2008} provided a Cepheid based distance modulus of 24.924 $\pm$ 0.042, and \citet{Bhardwaj2016A} derived a nearly identical distance modulus of 24.92 $\pm$ 0.08 based on new calibrations of the near-IR Wesenheit relations from Galactic and Large Magellanic Cloud (LMC) Cepheids. 

The consistency of these results was emphasized by the work of \citet{Lee2021} who found agreement within 0.01 mag for the distance moduli derived from Cepheids, the NIR TRGB, and the J-Branch Asymptotic Giant Branch (JAGB).  

\citet{Lee2021} measured an I-band TRGB-based distance modulus that was marginally closer in the mean, but still consistent within its uncertainties. Here, we provide an independent measure of the distance to WLM using new data, leveraging the \textit{Gaia}-consistent period-Wesenheit-metallicity relation \citep[PWZ,][]{Madore1982} in HST \citep[following ][]{Savino2022}.

\citet{Sarajedini2023} identified a statistically significant population of 90 RR Lyrae stars (76 ab-type) in the archival HST observations of WLM which are also used in this analysis.  He derived a mean metallicity of $\langle\textrm{[Fe/H]}\rangle = -1.74 \pm 0.02$ for the RR Lyrae stars (see discussion in \autoref{sec:metallicities}), lower than the population average of $\langle \textrm{[Fe/H]} \rangle = -1.28 \pm 0.02$ \citep{Leaman2013}. An RR Lyrae-based distance for WLM could not be simultaneously calculated, as a result of the degenerate relationship between the distance and metallicity determinations. The SFH from these HST observations has been presented in \citet{Albers2019}.
A new SFH from the JWST ERS observations of WLM has been presented by \citet{McQuinn2024}, as well as a spatially-resolved radial SFH presented by \citet{Cohen2025}. \citet{McQuinn2024} find that WLM experienced early star formation, followed by an extended ($\sim3$Gyr) post-reionization pause. After which, it experienced a re-ignition in star formation activity, which continues to present.

The structure of this report is as follows: in \autoref{sec:data} we describe the HST and JWST data used in this study, as well as the process of obtaining the correct data products for short-period variable work from the JWST archive. In \autoref{sec:HST}, we outline the identification, periodicity measurement, and MCMC template-fitting methodology \citep[from][]{Savino2022} for the HST data. In \autoref{sec:JWST}, we describe the same for JWST, specifically highlighting the new approach required for the shorter-baseline data, compared to HST. This section directly compares the identification results in HST and JWST, and calculates the recovery probabilities. Additionally, this section describes potential alternative applications of the short-baseline data for short-period variables, when MCMC fitting is unideal. Next, \autoref{sec:hstdistance} covers our PWZ-fit distance measurement using the HST data, including a discussion of various treatments for population metallicities. In \autoref{sec:jwstpwz}, we show the process for fitting the PWZ to the JWST bands, using the HST-derived distance as an anchor. In \autoref{sec:rrlpops} and \autoref{sec:discussion}, we describe our final RR Lyrae populations in HST and JWST, and discuss the final PWZ fits, including a comparison to alternative WLM distance measurements. Finally \autoref{sec:conclsions} outlines our conclusions, and a description of relevant future work.

\section{Data}\label{sec:data}
\begin{figure}
    \includegraphics[width=0.47\textwidth]{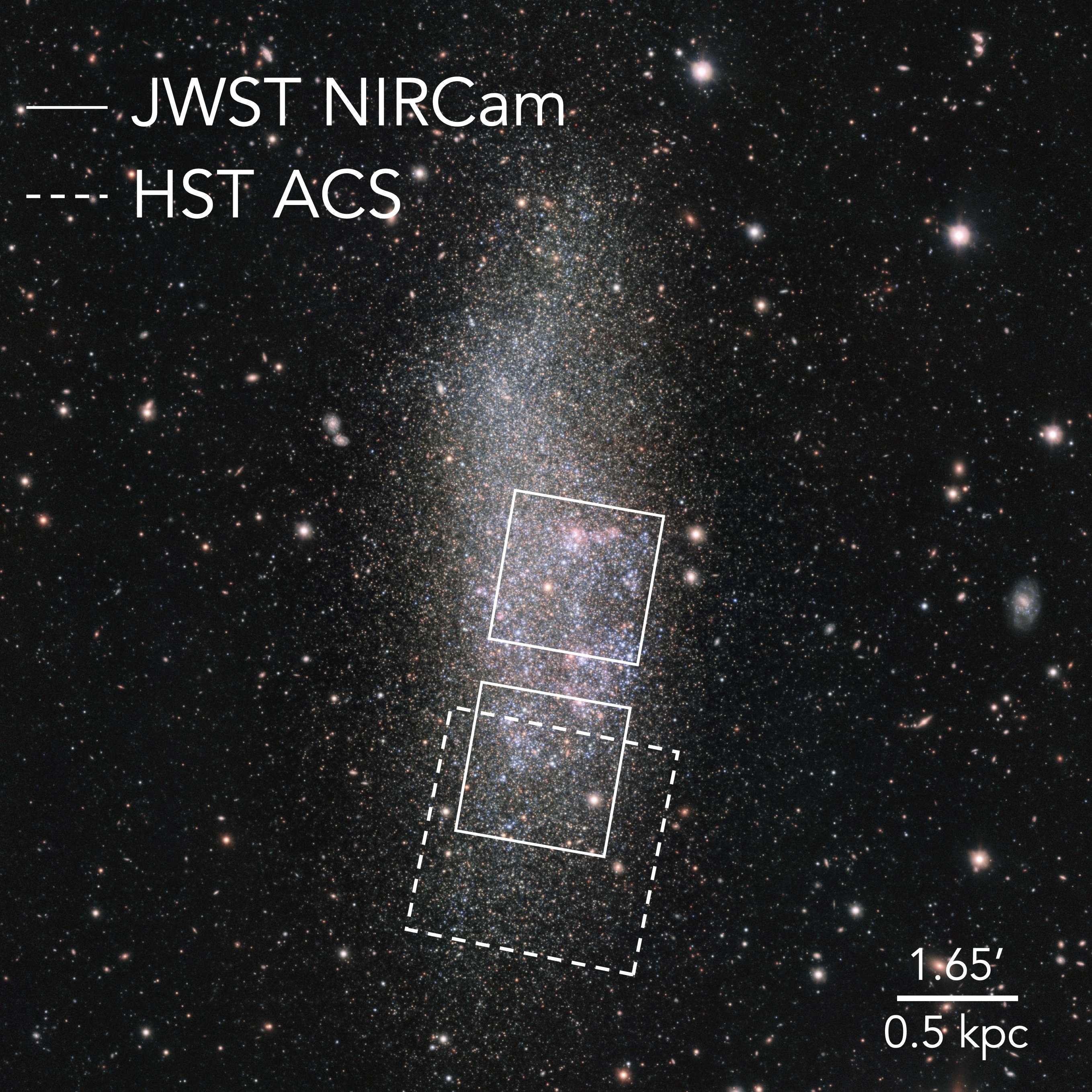}
    \caption{Ground-based optical image of WLM (image credit: ESO, VST/Omegacam Local Group Survey) with footprints of the NIRCam (solid), and HST ACS (dashed) fields overlaid. The outer NIRCam FOV overlaps with deep ACS F475W and F814W imaging. This overlap region of the deep HST and JWST observations enables a direct comparison of the RR Lyrae stars in both observations. Image adapted from Fig. 1 of \cite{McQuinn2024}.}
    \label{fig:fields}
\end{figure}

This paper makes use of archival data from both the HST and JWST archives, the field(s) for each are over-plotted with an image of WLM in \autoref{fig:fields}.

\subsection{HST Observations}
An ``inner'' field of WLM was observed with the HST ACS \citep[][]{Ford1998} for 25 orbits as part of the HST-GO-13768 program (PI: D.\ Weisz). ACS/WFC imaging observations were taken with the F475W filter (27,360 s) and the F814W filter (34,050 s). The imaging was obtained in 2-orbit visits spread over three days in July 2015 to allow for adequate coverage of short period variable stars.  Each orbit was split into one F475W and one F814W image. The filters were observed in reverse order during the second orbit for maximal cadence coverage. 

Point-spread function (PSF) photometry was performed using the \texttt{DOLPHOT} photometry package, which contains an ACS-specific module \citep{Dolphin2000, Dolphin2016}. Filtered photometric catalogs of the sources are created using the same culling parameters as in \cite{Savino2022}, originally outlined in \cite{Williams2014}. In particular, this includes a signal-to-noise (SNR) cutoff $\ge 4$, a sharpness squared cutoff $\le 0.2$, and a crowding cutoff $\le 2.25$.

\subsection{JWST Observations}
A ``central'' and an ``outer'' region of WLM were observed simultaneously with the JWST NIRCam instrument \citep[using the NIRCam A and B modules, respectively, see][]{Rieke2023} as part of the JWST Resolved Stellar Populations Early Release Science Program JWST-ERS-1334 \citep{Weisz2023}.
The NIRCam imaging totaled 30,492 s in the F090W filter and 23,707 s in the F150W filter.  The ``outer'' region has a significant overlap with  the area of the archival HST/ACS observations. Because the NIRCam imaging mode requires all dithers to be complete before changing filters, the total exposure time for each filter was observed continuously. 

\subsubsection{JWST Data Preprocessing} \label{sec:TSO}

In order to obtain the necessary time sampling for RR Lyrae from the JWST imaging, we begin by using the time-series pipeline to extract the individual integration data from the ERS observations, which were taken with the MEDIUM8 readout pattern with 8 groups, resulting in a cadence of 14 minutes per integration. This leaves us with 9 integrations per dither in F090W and 7 integrations per dither in F150W, with 4 dithers in each.
The standard JWST calibration pipeline \citep{2020ASPC..527..583B, 2023zndo...6984365B} retains individual integration data in calibrated Stage 2 products only for dedicated time-series observations (TSOs) due to data volume constraints. However, for observations taken in other modes --- such as deep imaging of crowded extragalactic stellar fields, where subpixel dithers are necessary to improve PSF sampling --- it is nonetheless possible for users to recover individual integrations for time-series analyses via a local pipeline run. For interested users, we recommend STScI's documentation \footnote{\url{https://jwst-pipeline.readthedocs.io/en/latest/jwst/user_documentation/introduction.html}} on how to set up the JWST data reduction pipeline locally, as well as the NIRCam documentation\footnote{\url{https://jwst-docs.stsci.edu/jwst-science-calibration-pipeline/key-differences-for-jwst-time-series-observations}} outlining TSO modes.

We began by retrieving the Stage 1 \texttt{*\_rateints} data products from MAST. The \texttt{*\_rateints} files are 3D data cubes of shape $ncols \times nrows \times nints$, where each slice along the $nints$ axis is a countrate image for the given integration.
We then processed these data with the Stage 2 imaging pipeline\footnote{\url{https://jwst-pipeline.readthedocs.io/en/latest/jwst/pipeline/calwebb\_image2.html}} (\texttt{calwebb\_image2} or \texttt{jwst.pipeline.Image2Pipeline}) to produce fully calibrated \texttt{*\_calints} cubes, which are the 3D equivalent of the 2D \texttt{*\_cal} files that our photometry software operates on.
As these are the products of interest for our analysis, we did not proceed to the Stage 3 pipeline step, which produces aperture photometry; however, such products could be of potential interest for less-crowded fields.

Next, we split each \texttt{*\_calints} cube into a series of individual ``pseudo-cal" 2D images. 
While this step is not strictly necessary, and indeed may be cumbersome at large data volumes, we chose to do so here for compatibility with our existing photometry routine, \texttt{DOLPHOT}.
We used the accompanying table of integration times (FITS extension ``INT\_TIMES") to assign correct \mbox{timestamps} to each pseudo-cal header.

From here, point-spread function (PSF) photometry was performed using the \texttt{DOLPHOT} photometry package, which contains a NIRCam-specific module \citep{Dolphin2000, Dolphin2016, Weisz2024}. Filtered photometric catalogs of the point sources are created using the purity-oriented selection from \cite{Weisz2024}. In particular, we establish an SNR cutoff $\ge 4$, a sharpness squared cutoff $\le 0.01$, and a crowding cutoff $\le 0.5$.

\section{HST Variable Star Identification and Fitting Methodology}\label{sec:HST}
The process for identifying variable stars in the catalog, isolating the RR Lyrae, and obtaining the parameters of variability has already been established by previous studies \citep[][]{Savino2022}. Here, in \autoref{sec:method-CI}, we describe the variable star candidate identification methodology. In \autoref{sec:HSTperiodicitymeasurment}, we elaborate on a more robust periodicity-measurement algorithm. Finally, \autoref{sec:HSTmcmctempfit} describes the MCMC template-fitting technique. We outline the entire analysis on the HST data first, to serve as a point of comparison for the JWST data in \autoref{sec:JWST}.

\subsection{Candidate Identification}\label{sec:method-CI}
 The resultant photometry catalogs are split into individual epochs. An epoch-to-epoch comparison is done \citep[as in][]{Dolphin2001, Dolphin2004} to pull out variable star candidates based on a number of photometric and variability assessments, enumerated below. In order to assure we obtain reasonably robust photometry, we only search for stars which appear in at least a minimum number of images. In the HST data, there are 24 images in F475W and 26 in F814W. A given star needed to be detected in at least two-thirds of the images (a total of 33). This is to avoid false positive identifications due to a small number of anomalous epochs.

In order to identify variability in our candidate set we look at three factors:

\begin{enumerate}
    \item To test for variability in the light curve, we require that the overall RMS scatter of the magnitude measurements for each star is greater than a threshold value. The RMS is a combined value for both bands.
    For HST, the RMS threshold is set to 0.1 mag, as in \cite{Savino2022}. This is above the average RMS in the photometry for the non-variable sources.
    
    \item To test that the variability is statistically significant, the reduced $\chi^2$ of each light curve is tested three times. For each test, we calculate the reduced $\chi^2$ for the individual-epoch data with respect to the mean magnitude, and require that it be \textit{greater} than a given threshold value, as a truly variable light curve will have measurements with statistically significant differences from the mean. For the first-pass, we test the full light curves. The cutoff threshold is set to a value of 2.6, based on the distribution of reduced $\chi^2$ values over the whole photometric catalog.

    The second-pass test is conducted on the light curves with the single most discrepant data point removed from each, to avoid interference by one-off contaminants (e.g., a cosmic ray). The threshold for the second-pass is the same as for the first.

    For the third-pass, we clip 1/3 of the most discrepant data points in order to eliminate non-variable sources with several bad observations. The threshold for this test is set to 0.5 for HST.

    \item To test that any observed variability is also periodic, the periodicity metric, $\Theta$, is computed according to the Lafler-Kinman (LK) phase dispersion minimization (PDM) algorithm \citep{Lafler1965} for test periods ranging from 0.25 to 3 days.

    \begin{equation}
        \Theta=\frac{\sum_{i=1}^N\left(m_i-m_{i+1}\right)^2}{\sum_{i=1}^N\left(m_i-\bar{m}\right)^2},
    \end{equation}

    For two filters, the individual metrics are combined as in \cite{McQuinn2015}. By definition, $\Theta$ is lower for a better-fitting guess in period. Therefore, the minimum $\Theta$ for the test periods must be less than 0.8 for the source to be considered sufficiently periodic for further consideration.
    
\end{enumerate}

\begin{figure}
    \centering
    \includegraphics[width=\linewidth]{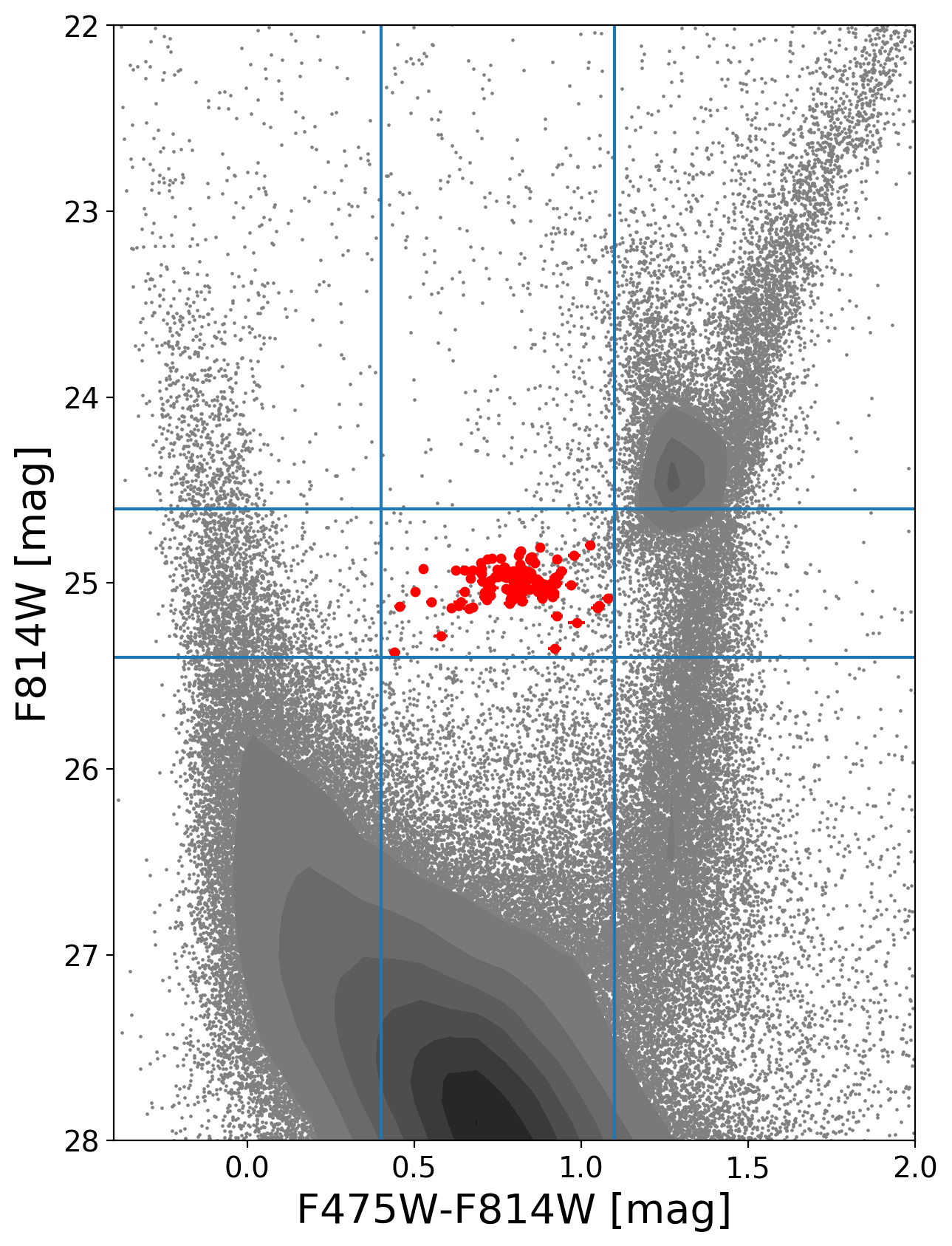}
    \caption{A CMD of our HST observations centered on the horizontal branch. The RR Lyrae candidates (red) are plotted over the full photometric catalog (grayscale). Uncertainties on the RR Lyrae are shown, but are too small to see for most sources. The most dense regions of the photometric catalog are indicated by contours, for viewing ease. The other sources are shown as points (lightest gray). The color-magnitude cuts are shown as blue lines.}
    \label{fig:hstcmd}
\end{figure}

If a given source passes our photometry, variability, and periodicity requirements, it is labeled a variable candidate and moved on to the next steps. 
In general, we err on the side of completeness over purity with this initial identification, as our initial sample is more rigorously filtered at later steps. We find 902 initial variable source candidates in HST.
\autoref{fig:hstcmd} shows an HST CMD with the RR Lyrae candidates highlighted.

\subsection{Periodicity Measurement}\label{sec:HSTperiodicitymeasurment}
While the candidate identification step outputs a first-guess at the period, we opt to employ the more rigorous method from \cite{Saha2017} for our formal periodicity measurement. This is a hybrid period-identification method, optimized for multi-band light curves with sparse sampling, which simultaneously makes use of the previously-described LK algorithm, and a Lomb-Scargle algorithm \citep{Lomb1976, Scargle1982}. 

\begin{figure}
    \centering
    \includegraphics[width=\linewidth]{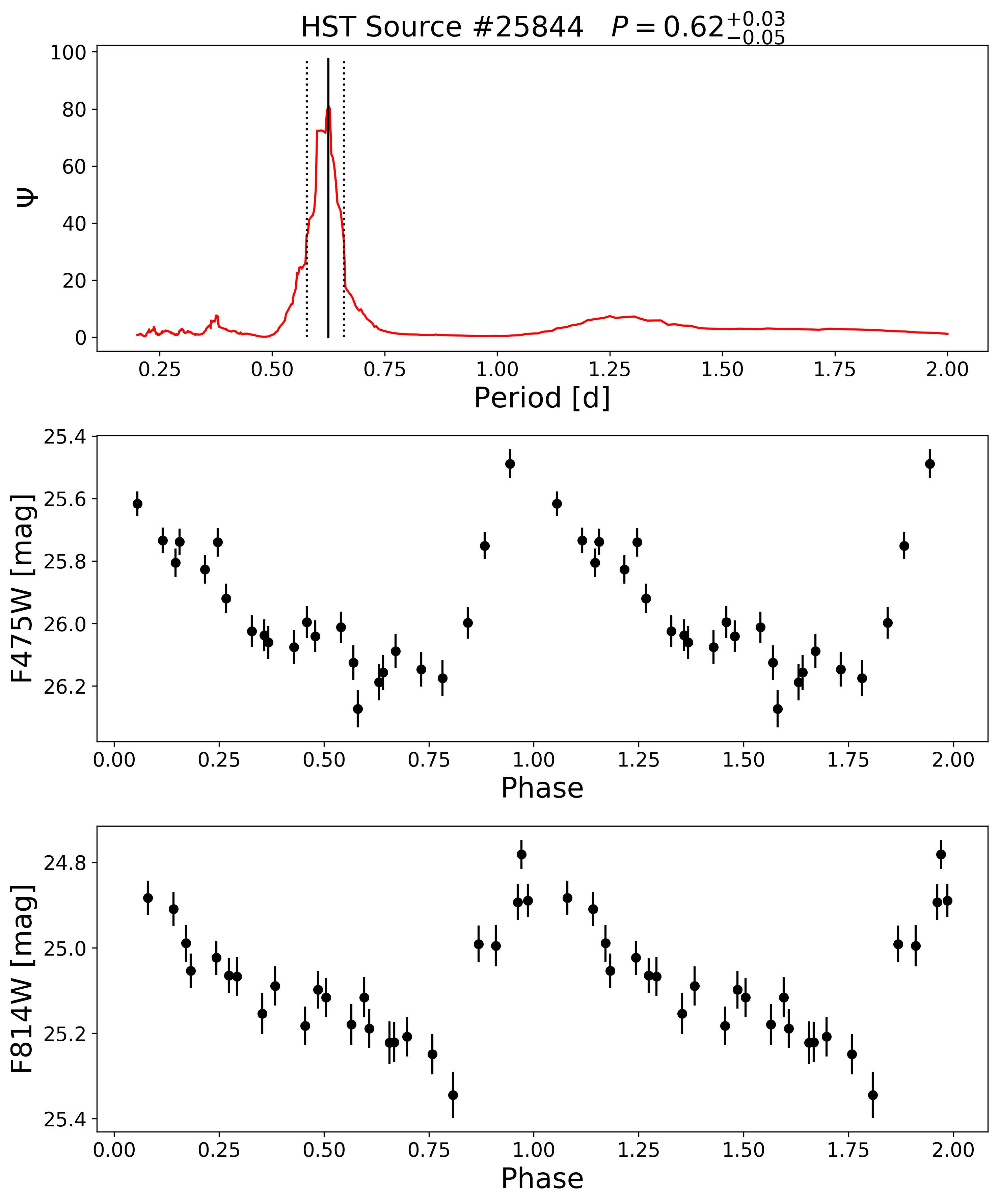}
    \caption{An example output from the periodicity measurement step. The top panel shows the power spectrum of the hybrid algorithm (red). The most likely period (solid black line) and errors (dotted black lines) are indicated. The middle and lower panels show the HST F475W and F814W data, phase-folded according to the most likely period.}
    \label{fig:periodogram}
\end{figure}

The Lomb-Scargle is a harmonic analysis (HA) algorithm, used in place of a Fourier-Series analysis. It relies on the amplitudes of fitted sine waves to glean the power at their various given frequencies. A full Fourier treatment would require computationally expensive resampling of the data, which can induce nonphysical structure into the overall power spectrum. Making use of a combined PDM and HA algorithm reduces the impact of artifacts introduced by the sampling. An example power spectrum can be found in \autoref{fig:periodogram}. The most likely period output by the hybrid method is used as a first-guess input for our final template fitting step.

During the periodicity measurement step, we also inspect the CMD and isolate the variable candidates which lie within about 0.4 mag from the mean of the Horizontal Branch (HB), and approximate color of the instability strip. For the HST bands, the HB region ranges from 24.6 to 25.4 mag in F814W, and 0.4 to 1.1 in F475W-F814W color (shown in \autoref{fig:hstcmd}). From these simple color-magnitude cuts, we identify 120 HST RR Lyrae candidates.

\subsection{MCMC Template Fitting}\label{sec:HSTmcmctempfit}
In order to obtain a final fit of our time-series data, we use a Markov Chain Monte Carlo (MCMC) sampling approach. We obtain Fundamental Mode (RRab) and First-Overtone (RRc) pulsator templates for our photometric bands. For our HST data, we make use of existing empirical Johnson B and I templates from \citet{Monson2017}. The optimal template selected for each filter is the one which minimizes the sum of squared residuals on the fit using the \texttt{scipy.optimize.minimize}\footnote{https://docs.scipy.org/doc/scipy/reference/generated/scipy.optimize.minimize.html} function \citep{Virtanen2020}. To improve computation time, there is some limitation on which templates are tested based on period, where relatively short- or relatively long-period sources only test c- or ab-type templates, respectively.

\begin{figure}
    \centering
    \includegraphics[width=\linewidth]{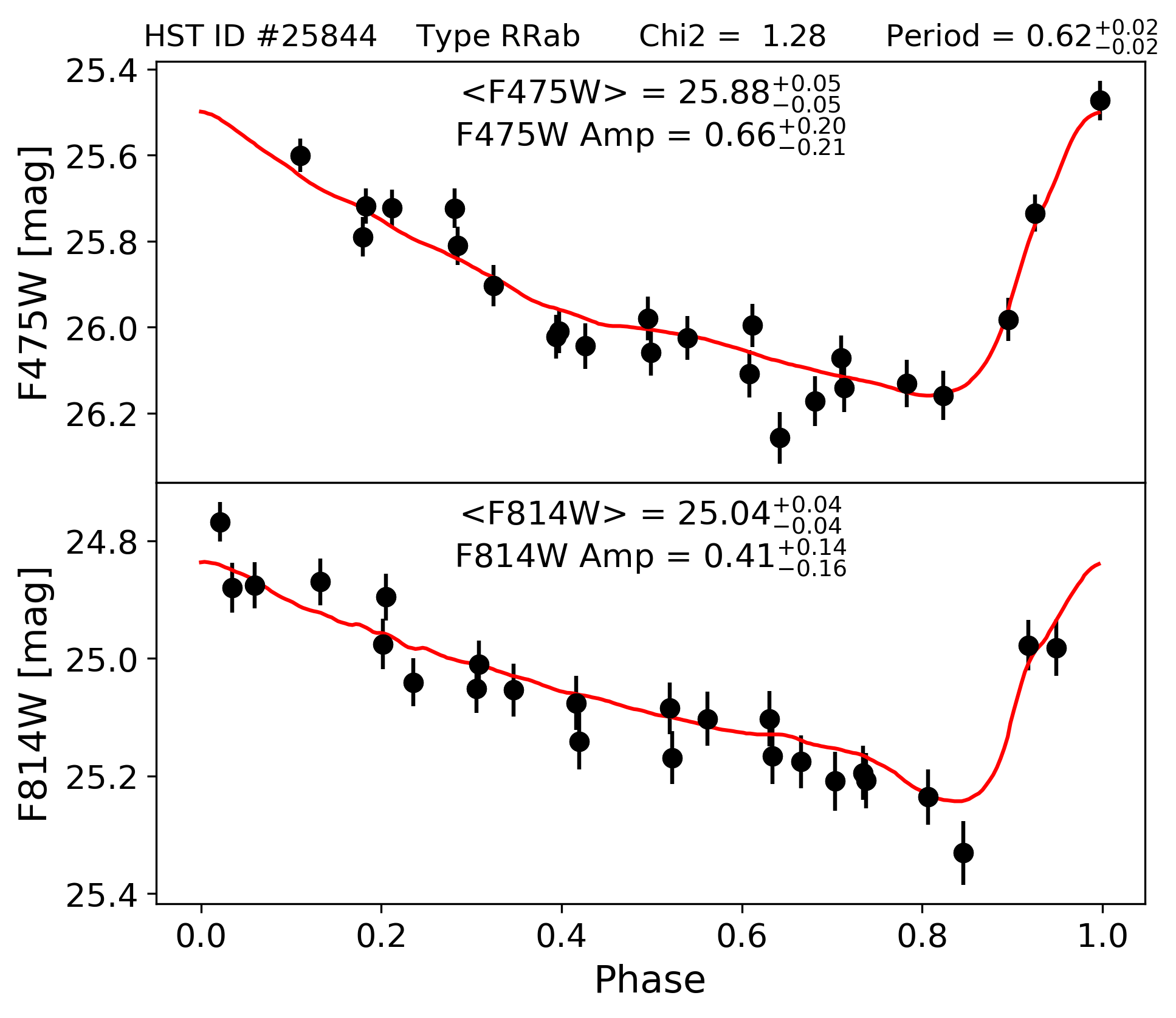}
    \caption{An example MCMC template fit to phase-folded HST F475W (top) and F814W (bottom) data. Observations (with uncertainties) are shown in black, while the MCMC fit is shown in red. The pulsation mode (RRab), $\chi^2$ on the fit, and period are shown above. The filter-wise mean magnitudes and amplitudes are written in their respective panels.}
    \label{fig:extemplatefit}
\end{figure}

Using a Gaussian likelihood function and broad, uninformative, uniform priors, we use the affine-invariant ensemble MCMC sampler \texttt{emcee} \citep{Foreman-Mackey2013} to sample the posterior probability distribution for each star. We define the convergence length of the MCMC chain as 50 times the autocorrelation length. In doing so, we identify constraints on the parameters of pulsation (period, band-dependent amplitudes, etc.) with uncertainties, and identify the pulsation mode (based on the period and best-fitting template). An example model fit can be found in \autoref{fig:extemplatefit}.

\begin{figure}
    \centering
    \includegraphics[width=\linewidth]{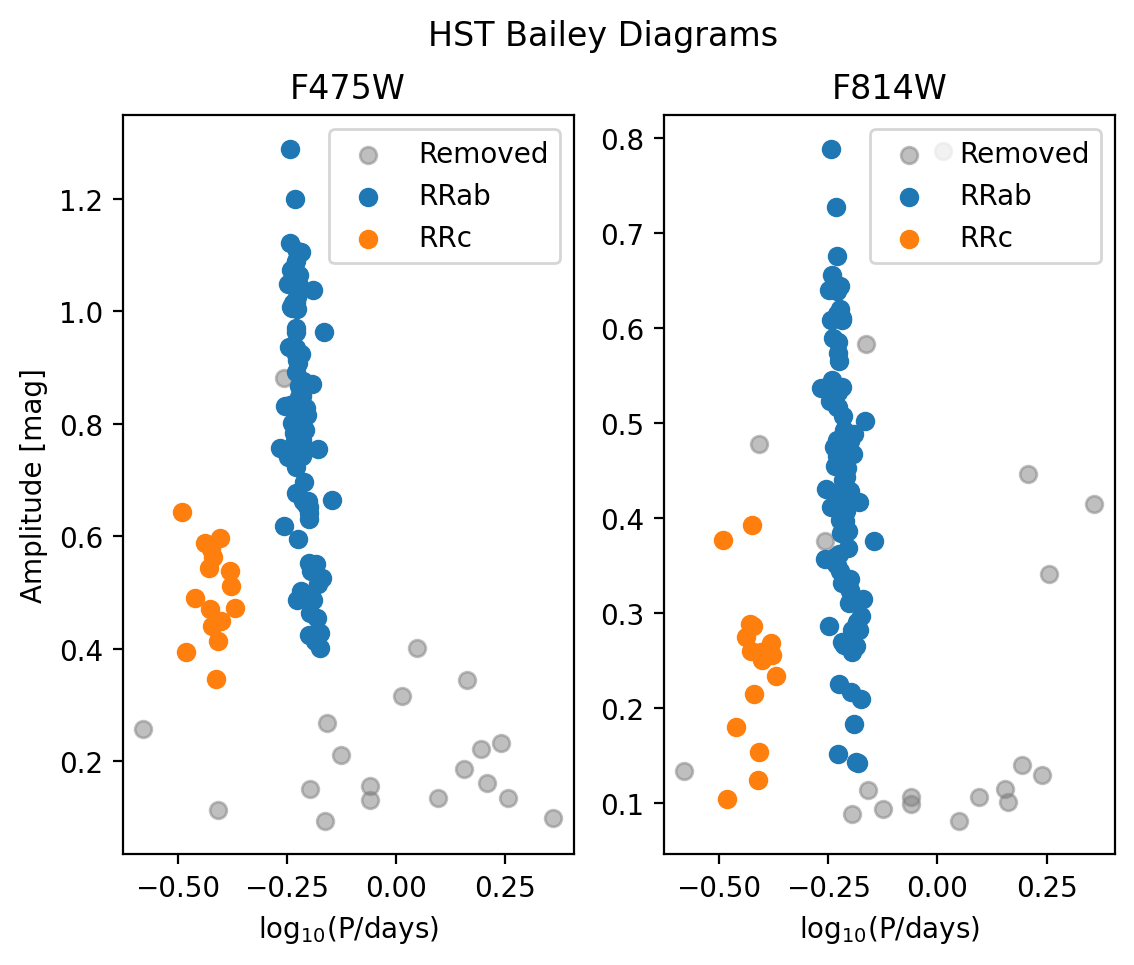}
    \caption{Bailey diagrams created from the HST RR Lyrae. Each point represents the best-fit period (x axis) and amplitude (y axis) output by the template-fitting MCMC for each source. The F475W data is in the left-hand panel, the F814W data in the right. Sources best-fit as RRab sources are shown in blue, while those best-fit as RRc sources are shown in orange. Variable sources which were removed from the data set upon visual inspection are in gray.}
    \label{fig:HSTbailey}
\end{figure}

From the model fit, we obtain best-fit periods, amplitudes, and mean magnitudes. We make use of model fits and posterior distributions, as well as diagnostic plots such as period-amplitude (``Bailey'') diagrams and CMDs to visually inspect and address irregular sources. A handful of sources 
which exhibit obvious and consequential aliasing are rerun with a more constrained prior on the period. In addition, sources which have obviously been mis-typed in the template selection step (based on their periods and locations on the Bailey diagrams) are rerun with the alternate type enforced. Finally, we remove any remaining non-RR Lyrae, as well as two sources with MCMC-derived period uncertainties which were too large for use in the PWZ fitting. Our HST Bailey diagrams can be found in \autoref{fig:HSTbailey}. We ultimately identify 101 \textit{bona fide} RR Lyrae in the HST data, as compared to the 90 originally identified in \citet{Sarajedini2023}. 
Source coordinates to cross-match against (and uncertainties on the derived values) are not reported in \citet{Sarajedini2023}, so a deeper source-by-source comparison of the fitting code proves problematic at this time.
The HST epoch photometry and characteristic parameters (RA and Dec coordinates and template-fitting outputs) for these final RR Lyrae can be found in \autoref{sec:tables}, \autoref{tab:HSTepoch} and \autoref{tab:HSTfinalfits}, respectively.

\section{JWST Variable Identification and Fitting Methodology} \label{sec:JWST}

As outlined in \autoref{sec:intro_VS}, there are several factors that make analyzing RR Lyrae somewhat more difficult in JWST than in HST. In addition, for our WLM JWST observations, the $\sim 15$ hr baseline is short enough that the majority of our sources do not have much more than one full cycle of data. A subset of these will not even have an entire cycle. This can pose problems in particular for the periodicity measurement and MCMC fitting steps, as they will be more influenced by spurious effects. 

The general analysis for identifying and fitting the RR Lyrae as described in the previous section remains largely the same for JWST. Here, we outline the additional considerations and analysis conducted specifically for the short-baseline JWST data.

\subsection{Candidate Identification}
\begin{figure}
    \centering
    \includegraphics[width=\linewidth]{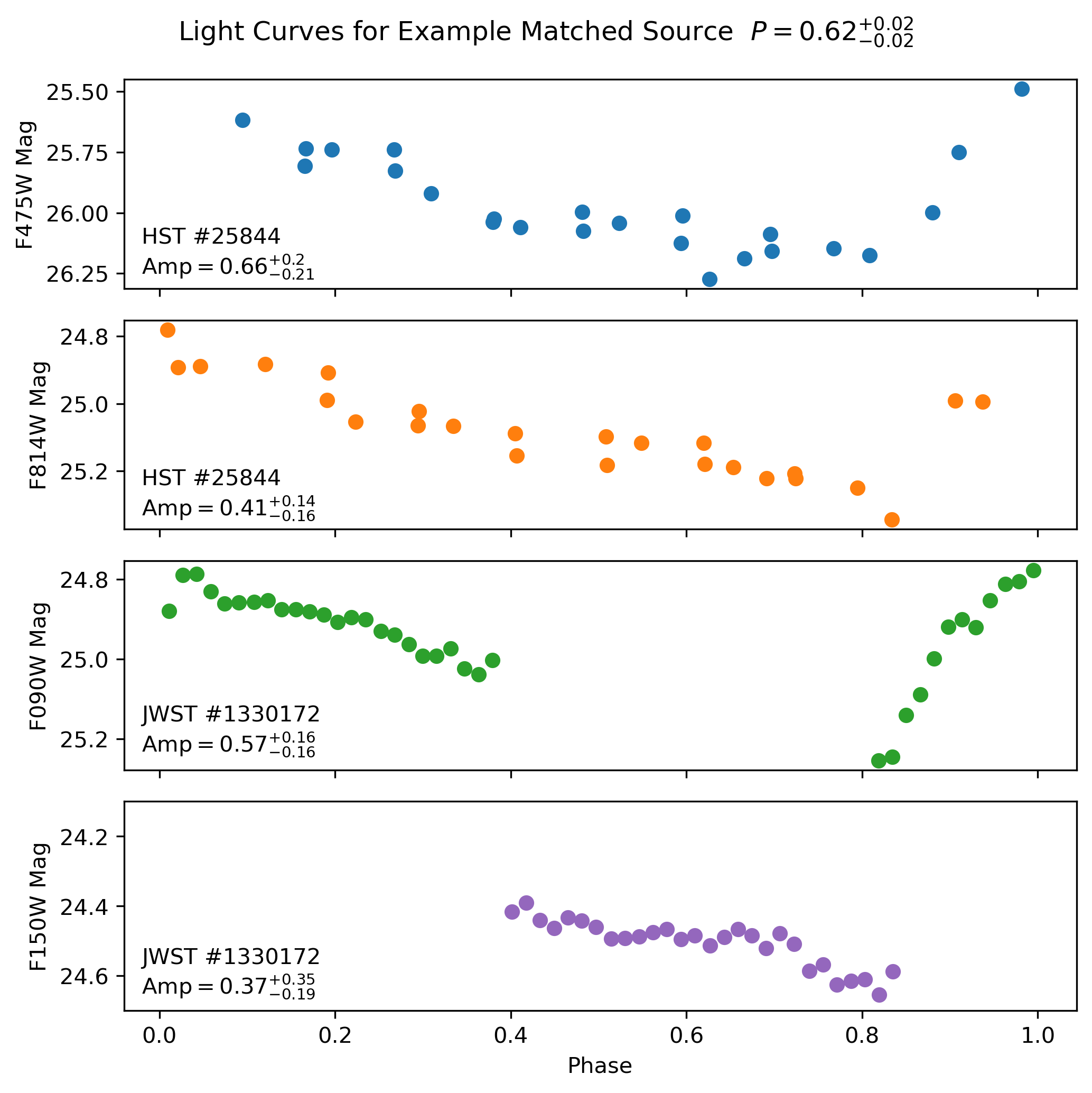}
    \caption{The HST and JWST light curves for the same example source shown in \autoref{fig:periodogram}. The top two panels show the HST F475W (blue) and F814W (orange) data. The bottom two panels show the JWST F090W (green), and the F150W (purple) data. The curves have been phase-folded according to the HST-derived period ($\sim0.62$ days), and phase-shifted to roughly align the extrema for viewing ease. Note the similarity in light curve shape, as well as the differences in cadence and coverage due to observation methodology.}
    \label{fig:goodlightcurves}
\end{figure}

To mitigate the effect of the smaller amplitudes in the JWST bands, we conduct an analysis of cutoff parameter values (outlined in \autoref{sec:method-CI}) which allow us to recover the greatest number of HST-identified RR Lyrae in the overlap region. In principle, the identification code (and color-magnitude cuts to isolate the horizontal branch) ought to find the same RR Lyrae in the JWST data that are found with HST. However, in reality, this recovery is imperfect. The severity of this loss is, understandably, affected by our choice of cutoff. For reference, example HST and JWST light curves for a successfully recovered RR Lyrae source can be found in \autoref{fig:goodlightcurves}. The similarity in general shape can be seen, and the differences due to observation methodology are apparent. On the flip side, we also do not want to misidentify stars in JWST which are not found to be variable in HST. In general, the photometry-based quality cuts are largely the same in JWST as they are in HST.

To check that the observed variability is sufficiently large, the RMS of the data is compared to a threshold value, as described in \autoref{sec:method-CI}, item \#1. For JWST, this value is determined via a comparison of the average RMS for \textit{all} stars in both data sets, and scaled accordingly, resulting in a cutoff of 0.04 mag. We secondarily assess how reasonable our threshold value is by looking for false positive variables (see \autoref{sec:falsepositives}).

\begin{figure*}
    \centering
    \includegraphics[width=.7\linewidth]{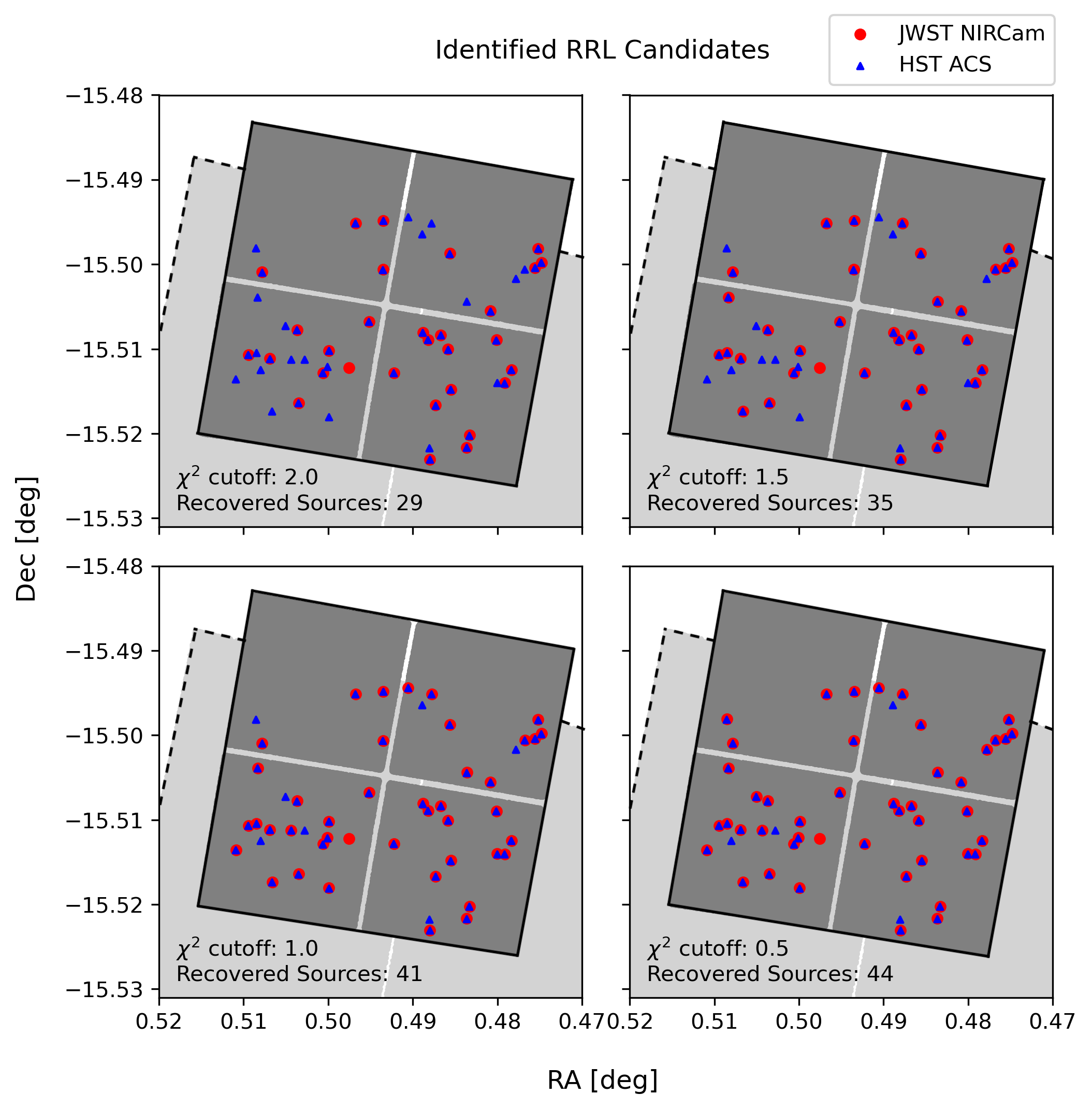}
    \caption{A selection of JWST (red circles) candidate data sets compared to the HST (blue triangles) candidate data set, for four different final $\chi^2$ thresholds on the JWST data. In each panel, the set of HST sources is the same, while the set of JWST sources refers to those identified with the associated given $\chi^2$ cutoff. ``Recovered Sources" refer to those which are found in both HST and JWST. Data are plotted in sky coordinates, with the JWST (dark gray) and HST (light gray) footprints shown. Only stars in the overlap region are pictured. False positives are locations with only a red data point, and false negatives are locations with only a blue data point, while correctly matched stars will show both. As the chi-squared threshold decreases, more stars are identified in the JWST data set, without an increase in false positive sources. One of the remaining HST sources that is not identified in JWST is ultimately removed from the variability identification testing, as it is eliminated in JWST for early quality cuts, not lack of variability.}
    \label{fig:chi2check}
\end{figure*}

In order to ensure that the observed variability is authentic, the reduced $\chi^2$ of the variance is calculated three times. The first- and second-pass cutoffs are set to 1.0, based again on the distributions. For the third-pass threshold, we check a variety of reduced $\chi^2$ values to best maximize recovery of the HST candidates without identifying too many false positives. \autoref{fig:chi2check} shows the HST RR Lyrae in the overlap region, as well as the JWST RR Lyrae found with four different third-pass $\chi^2$ thresholds.

Unsurprisingly, we find that the recovery of the HST dataset is better as we decrease the minimum allowable $\chi^2$ (for the case when 1/3 of the data points are removed). In addition, we note that we only identify one candidate false positive source in JWST, for all the third-pass cutoffs tested. Further notes on this source can be found in \autoref{sec:falsepositives}.

Based on this analysis, we set the JWST third-pass threshold to 0.5, the same as for HST. Ultimately, we find 706 variable source candidates in the JWST data.

\begin{figure}
    \centering
    \includegraphics[width=\linewidth]{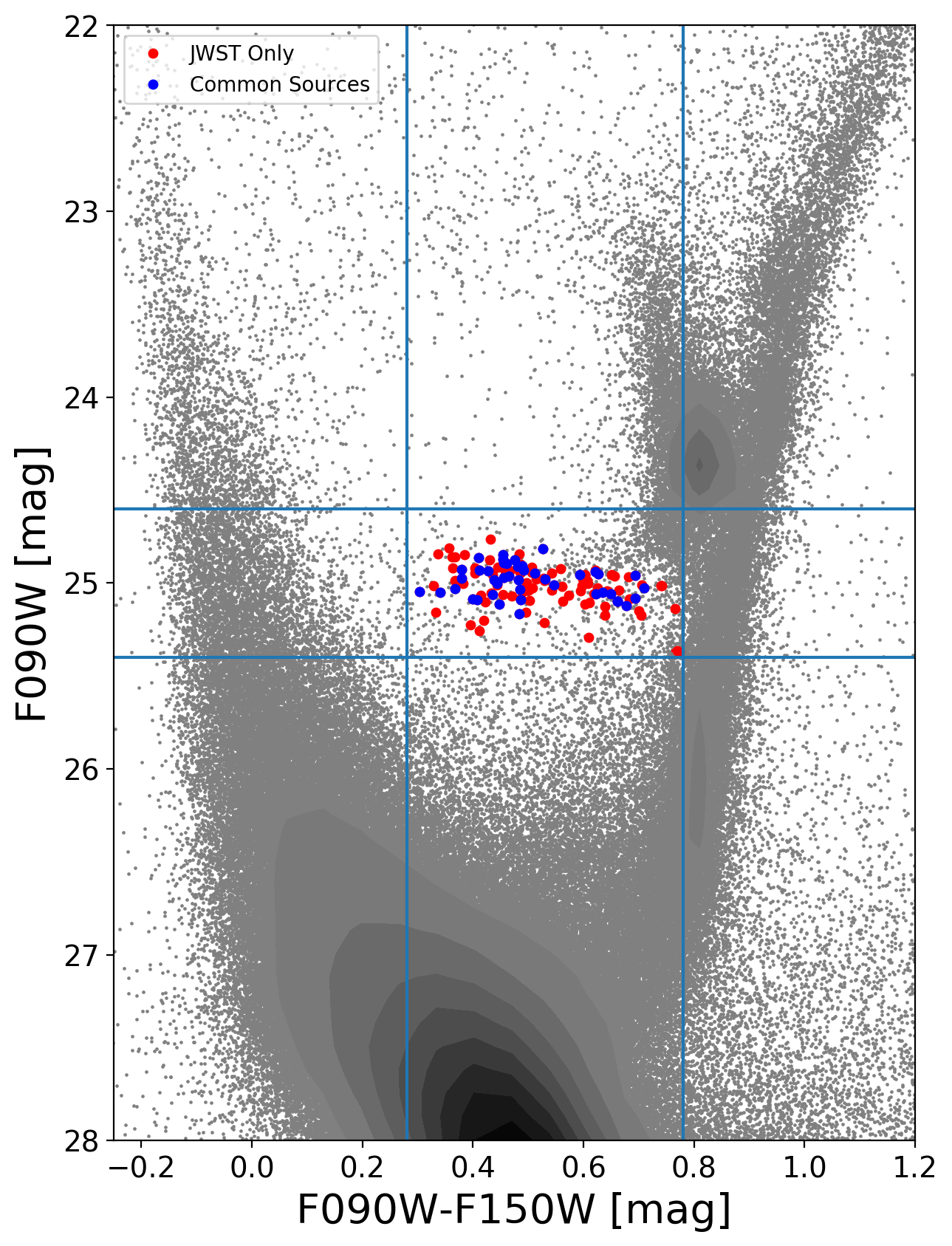}
    \caption{A CMD of our JWST observations centered on the horizontal branch. The 125 final RR Lyrae candidates are plotted over the full photometric catalog (grayscale). Sources which are found only in JWST are in red, while sources found in both HST and JWST are shown in blue. Uncertainties on the RR Lyrae are shown, but are too small to see for many sources. The most dense regions of the photometric catalog are indicated by contours, for viewing ease. The other sources are shown as points (lightest gray). The color-magnitude cuts are shown as blue lines. As compared to the HST CMD (\autoref{fig:hstcmd}), there appears to be greater scatter about the HB, most likely a result of taking the means of incomplete lightcurves for long-period RRab sources.}
    \label{fig:jwstcmd}
\end{figure}

Of these, 137 are flagged as potential RR Lyrae, based on their proximity to the HB. 12 of the 137 are removed during the attempted period-identification step (\autoref{sec:jwstperiodicityMCMC}) based on location on the CMD, and visual inspection of the output periodograms and light curves. This leaves a final candidate RR Lyrae population of 125. These are plotted, along with the general catalog, on the CMD in \autoref{fig:jwstcmd}. Similar to that described in \autoref{sec:HSTperiodicitymeasurment}, the HB region ranges from 24.6 to 25.4 mag in F090W, and from 0.28 to 0.78 in F090W-F150W color. The similarity in magnitude cutoffs between HST and JWST is a result of the similarity in the span of the HST F814W and JWST F090W bands (see \autoref{fig:filters}). The JWST epoch photometry for the 125 RR Lyrae candidates can be found in \autoref{sec:tables}, \autoref{tab:JWSTepoch}.

\subsubsection{Parameter Space of Unmatched Stars}\label{sec:parameterspace}

\begin{figure}
    \centering
    \includegraphics[width=\linewidth]{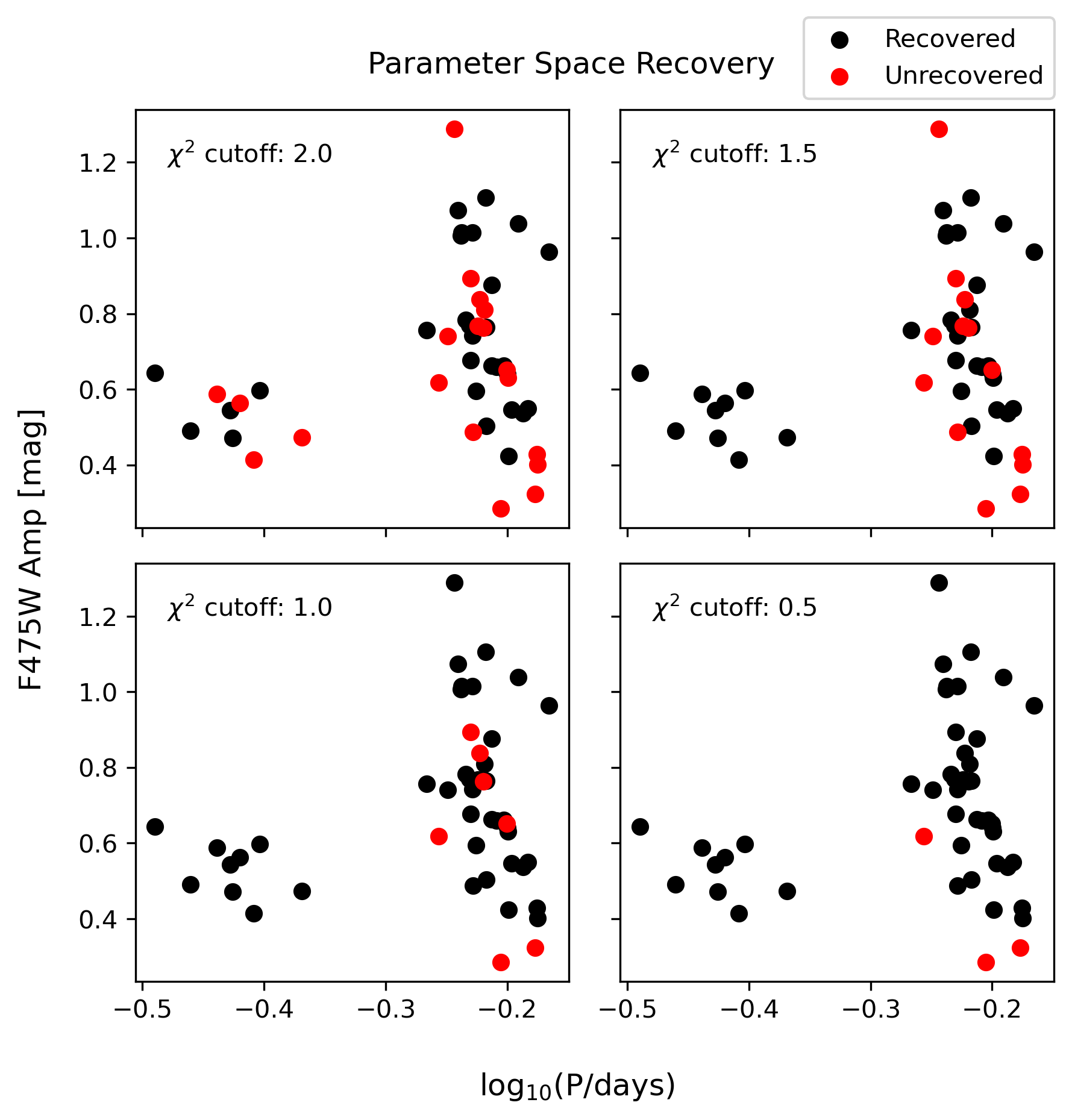}
    \caption{A selection of 4 Bailey diagrams, created from the HST MCMC F475W output fits, highlighting the parameter space most easily missed in the JWST data. Point color indicates if a given source was found by the JWST variability identification for 4 different final $\chi^2$ thresholds. In the lower right panel (threshold 0.5) only the final three false negative sources remain unmatched.}
    \label{fig:chi2bailey}
\end{figure}

For guidance on future RR Lyrae research using JWST, it is useful to identify possible selection effects and patterns. In general, we expect to disproportionately miss stars that are particularly low amplitude, or long period (especially given the relatively short time baseline of the WLM observations we are currently working with). To illustrate, \autoref{fig:chi2bailey} shows the WLM Bailey diagram, based on the HST data and, for various values of the third-pass $\chi^2$ threshold, indicates which stars are correctly identified as RR Lyrae in the JWST data set. Doing so allows us to look for patterns in the joint period-amplitude space, which is ultimately more illustrative than assessing either parameter in isolation.

We find that, for high third-pass $\chi^2$ cutoffs, the region of the Bailey diagram corresponding to long-period stars is most likely missed, with high-amplitude and short-period sources perhaps being preferentially found. As we decrease the cutoff threshold, it appears that we next pick up short-period RRc stars (in other words, stars with a greater number of observed cycles), and the impact of amplitude is reduced. Surprisingly, it appears that mid-amplitude RRab stars are some of the last to be entirely recovered. This may be explained by the greater number density of stars in this region of the Bailey diagram. It should be highlighted that, for sufficiently low thresholds, there is no part of the parameter space that is entirely missed. A description of how to extend this analysis can be found in \autoref{sec:futurework}.

\subsubsection{Inspection of the False Positive and False Negatives}\label{sec:falsepositives}

\begin{figure}
    \centering
    \includegraphics[width=\linewidth]{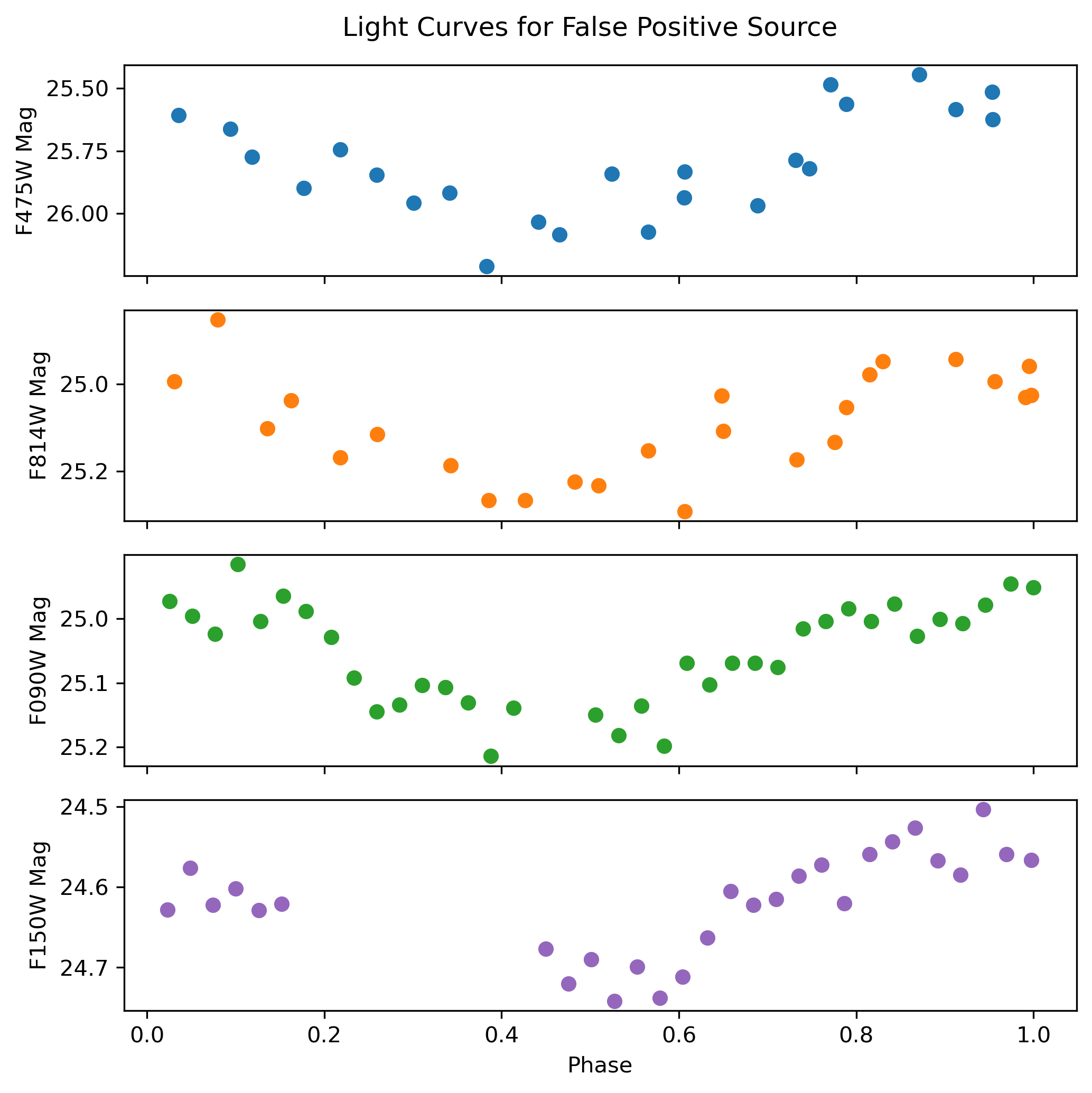}
    \caption{The HST and JWST light curves for the false positive source. The top two panels show the HST F475W (blue) and F814W (orange) data. The bottom two panels show the JWST F090W (green), and the F150W (purple) data. The curves have been phase-folded according to the best-guess period output by the Lafler-Kinman algorithm ($\sim0.38$ days), when run on the HST data, strictly for viewing ease. Robust peak-to-peak amplitudes are necessarily not reported, as this source was not included in the HST MCMC template-fitting.}
    \label{fig:fplightcurves}
\end{figure}

Until now, we have taken for granted that the identification of false positive variable sources to be an indication of an error in the processing and variability analysis of the JWST data. However, visual inspection of the single false positive source in the JWST data shows that it appears to be genuinely variable. \autoref{fig:fplightcurves} shows the source light curve, the third and fourth panels are the JWST data.

This variability indicates that this star may not be an actual false positive. Re-running the variable identification step in HST with lower $\chi^2$ and RMS thresholds does not recover this star. 
We identify that star in the HST observations through a positional match  
and determine that it \textit{would} be identified as variable in HST (see the \autoref{fig:fplightcurves}, panels 3 and 4), but it first fails the photometry quality check (specifically, the crowding is too high).  Thus, the higher angular resolution of the JWST has allowed the recovery of this star, which is not a false positive, when HST did not.

There is one source that is identified in HST as variable, but fails the initial quality cuts in the JWST photometry and, as such, is never analyzed for variability in those filters. We have elected to retain this source in \autoref{fig:chi2check} for transparency, but do not consider it in our analysis of the the efficacy of the variability identification in the JWST bands.

In addition, there are three relevant false negative stars in the overlap region—identified as RR Lyrae in HST—which are not picked up by the variability identification step for JWST. Visual inspection of the HST light curves confirms that these sources appear to be variable. 
From positional matches, we find that all three of these false negative sources fail the RMS threshold check in the JWST observations (\autoref{sec:method-CI}, item \#1), with RMS values ranging between  $0.033$ and $0.037$. However, decreasing the RMS cutoff value in JWST to include these stars results in a significant increase in the number of false positive sources. In order to prioritize the purity of our sample, we leave the JWST RMS threshold at the previously stated 0.04.

\subsubsection{Calculation of Observation Probabilities}\label{sec:probabilities}

The overlapping data affords us the ability to calculate estimates of the probabilities of detection in HST and JWST, as well as the total number of observable RR Lyrae in the shared region. For this, we implement a fairly simple approach. If $N_{RRL}$ is the total number of RR Lyrae in the overlap region, and $P_{HST}$ and $P_{JWST}$ are the probabilities a given RR Lyrae is detected by HST and JWST, respectively, then we can construct a set of equations anchored to our observations:
\begin{equation*}
    \begin{split}
    n_{HST \cap JWST} &= N_{RRL}P_{HST}P_{JWST} \\
    n_{HST\lnot JWST} &= N_{RRL}P_{HST}(1-P_{JWST}) \\
    n_{JWST\lnot HST} &= N_{RRL}(1-P_{HST})P_{JWST} \\
    \end{split}
\end{equation*}
Where $n_{HST\cap JWST} = 44$, $n_{HST\lnot JWST} = 3$, and $n_{JWST\lnot HST} = 1$ are the number of RR Lyrae observed in \textit{both} datasets, in HST but not JWST, and in JWST but not HST, respectively. Plugging in the appropriate values and solving yields the expected maximum likelihood values for $N_{RRL}$, $P_{HST}$, and $P_{JWST}$. We also use an MCMC sampling technique with a multidimensional binomial likelihood distribution to estimate the associated uncertainties.
\begin{equation*}
    \begin{split}
    N_{RRL} &=  48\\
    P_{HST} &= 0.96 \pm 0.03\\
    P_{JWST} &= 0.92 \pm 0.04\\
    \end{split}
\end{equation*}
We note that the posterior distribution on $N_{RRL}$ is necessarily discrete and asymmetric (with a minimum of 48), meaning standard metrics to report uncertainty lack the same statistical relevance. In the posterior distribution, 84.5\% of samples correspond to $N_{RRL}=48$, 13.3\% of samples correspond to $49$, and the final $\sim$2\% have values greater than $49$, with the maximum value at 56.

These results indicate that our expected completeness is certainly high enough in order to measure the distance for present purposes. Future studies which aim to prioritize completeness of their RR Lyrae sample may require an alternative treatment.

It should be noted that the above probabilities are calculated for the respective observational baselines in our HST and JWST data. A discussion of a more thorough treatment of estimating completeness, taking into account varying baselines, is found in \autoref{sec:futurework}.

\subsection{Periodicity Measurement and MCMC Fitting}\label{sec:jwstperiodicityMCMC}

When attempting to measure the RR Lyrae periods from the JWST data, we encounter issues on the periodicity-measurement and template-fitting steps. Specifically, following the same prescriptions used on the HST data, we are unable to recover the correct periods for most of the sample. In \autoref{sec:appendix}, we test the potential impact of epoch-to-epoch zero-point offsets due to persistence in the NIRCam instrument, and find no significant effects.  The main challenge in working with these observations is the limited time baseline compared to the periods of these stars.  For a PWZ distance analysis, accurate periods with secure uncertainties are required.  However, as we discuss here, the short time baseline of these observations prevents the success of a standard analysis.

In general, the periodicity measurement and MCMC fitting are conducted as outlined in \autoref{sec:HSTperiodicitymeasurment} and \autoref{sec:HSTmcmctempfit}. For the MCMC fitting, we use similar templates from the standard I and H bands, which reasonably map to the F090W and F150W filters, respectively. The I-band templates again come from \cite{Monson2017}, while the H-band templates were created later by \cite{Braga2019}.

\begin{figure}
    \centering
    \includegraphics[width=.9\linewidth]{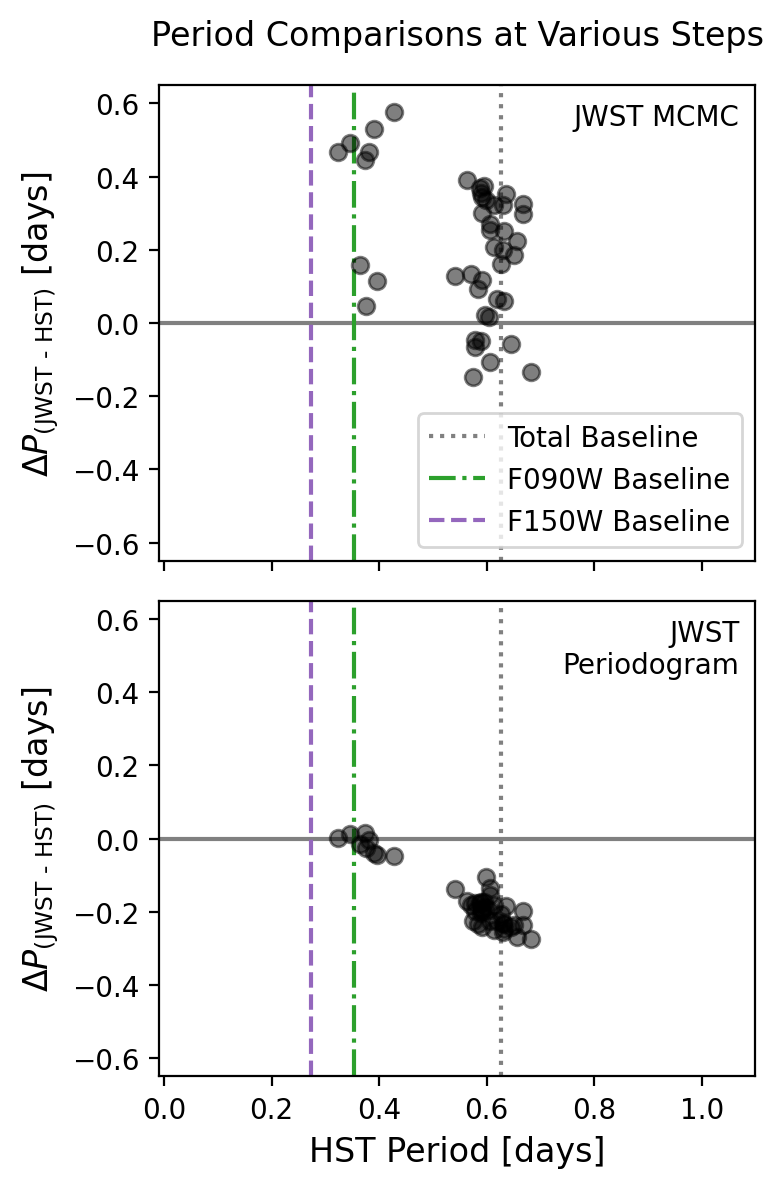}
    \caption{The two diagnostic scatter plots. The top plot shows the difference between the JWST MCMC output and the HST-derived period for each overlapping source, while the bottom plot shows the difference between the JWST periodogram output and the HST period. The solid horizontal lines represent the expected value $\Delta P =0$. The gray vertical lines indicate our JWST observational baseline of about 15 hours. The green and purple vertical lines represent the baselines for the individual observational baselines for the F090W and F150W bands, respectively. While the periodogram is able to obtain reasonable periods for the RRc sources, it systematically under-estimates the higher-period RRab sources. Similarly, the MCMC template fitting fails to identify the expected periods for both classes of RR Lyrae.}
    \label{fig:diagnostic}
\end{figure}

In order to assess the period-finding and classification in JWST, we again make use of the common stars in the overlap region. From these, we make two diagnostic scatter plots, shown in \autoref{fig:diagnostic}. The first shows the final MCMC period fits for the HST and JWST data sets, and the second compares the JWST periodogram output with the HST MCMC fits. Because the HST MCMC fits are sufficiently close to the true values, we can use them as a point of reference to determine where the error may be occurring in our JWST fitting.

For each of these plots, we would expect the data to sit nearby and randomly scattered about the $\Delta P =0$ line, as our methodology should recover the ``real" values. In the first, the JWST MCMC output clearly does not do so. We can clearly identify the regions of the HST MCMC axis where the RRab- and RRc-type sources sit, but this stratification does not occur in the JWST MCMC output space. There are two places in our methodology where this error could be introduced.

\subsubsection{Periodicity Measurement}
The first is the periodicity-measuring step, as represented by the lower plot in \autoref{fig:diagnostic}. In this plot, we see sources with relatively short periods are being well identified by the periodicity-measuring step. These are all RRc-type pulsators, based on the HST analysis output. For the other sources, they have periods broadly sitting around an estimated true value equal to our JWST observation duration. Meaning that we only have about a single cycle of data for the periodicity measuring step to work with. Additionally, the nature of the chosen JWST observing scheme means that each filter has an observational baseline about half the total for each source. While the power calculated for the periodogram is a combination of the algorithm outputs for each filter, they are ultimately fit independently, reducing the effective phase-space coverage. This further enforces that, for data where the filter observations are not inter-spaced, a single cycle is insufficient to effectively run this algorithm, given that only the short-period variables are being well recovered  \citep[see further discussion of aliasing in][]{Savino2022}.

Further assessment of the output from the \cite{Saha2017} method shows that for many of the long-period sources, the periodograms for the JWST data flatten out very abruptly at longer test periods, creating a broad range of possible periods with near-identical power. For many of the long-period RRabs, there is not a clear peak in the power spectrum before the flattened portion. Upon further inspection, we determined that this flattening occurs primarily because the Lafler-Kinman phase-dispersion minimization will assign equal power to any test period greater than the observational duration, because of the phase-folding step. For example, when 16 hrs of observations are phase-folded with a 16- or 20-hr period, the relative positions of the data points will be the same, yielding the same Lafler-Kinman periodicity metric. Further testing focused instead on the Lomb-Scargle piece of the period-finding.

We also assessed the impact this may have on the variability-finding step, as the Lafler-Kinman is used to estimate periodicity in the variable-source-finding step, and find no issue. The relatively modest number of false positive and false negative sources confirms this. In the case of the variability identification, we are simply searching for indication of some non-random trend in variability, which is still exhibited in the short-duration light curves, and is sufficiently captured by the Lafler-Kinman metric for our identifications scheme.

\begin{figure}
    \centering
    \includegraphics[width=\linewidth]{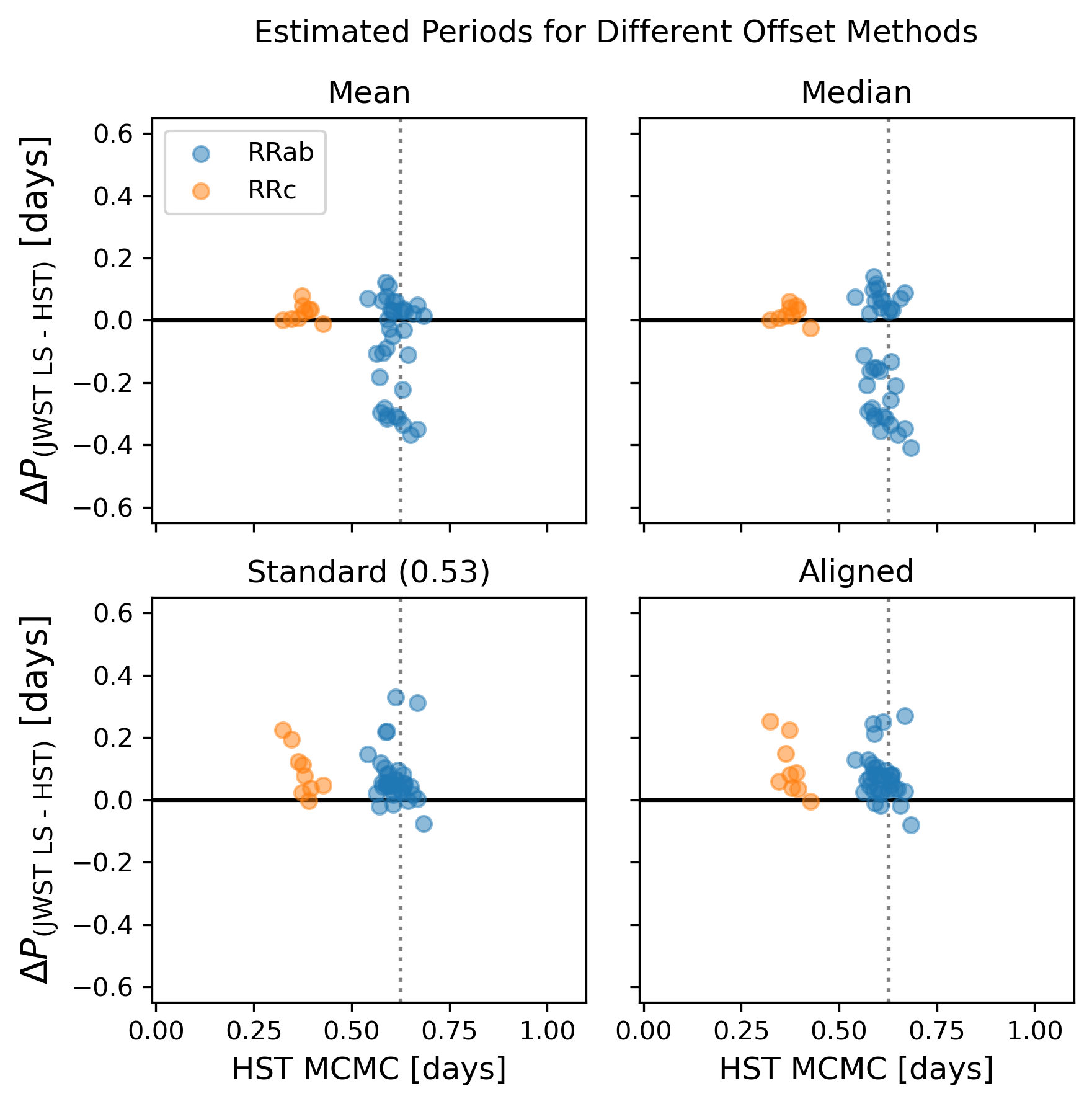}
    \caption{The HST-derived periods versus the period offsets from our various single-data-stream tests for four different combination methods. $\Delta P = 0$ is shown in black, and the JWST observational duration as gray dotted lines. Sources are colored according to their HST-derived classification (RRab, blue; RRc, orange). We find that the alignment procedure works best on a population-wide level, while avoiding the random guessing of a standard shift.}
    \label{fig:offset_test}
\end{figure}

In light of the filter-splitting issue, we attempted to find the period on both data sets simultaneously (using \textit{just} the Lomb-Scargle periodogram) as if they were a single-filter data stream. For this, we tested a number of methods for ``combining'' the light curves. \autoref{fig:offset_test} shows the updated period comparison plots for the four tested methods. These methods are: shifting one filter so the filter-wise means were equal (``mean''), shifting so the medians were equal (``median''), shifting by a pre-determined, fixed standard amount (``standard''), and aligning the light curves such that the magnitudes of the final point for the F090W and the first point in the F150W were equal (``aligned''). We determined the aligned method appears to be most effective across the board, though it does markedly worse than the mean or median methods on the C-type sources alone. The aligned method also prevents the need to make a good guess on a standard shift. Attempts to also adjust the amplitudes did not yield additional improvement. While imperfect, this method may be used
to obtain a reasonable first-pass period guess for the longer period variables with insufficient time baseline coverage.

\subsubsection{Template Fitting}
The second place the period estimation fails is the actual MCMC template fitting, which takes in the output of the periodicity algorithm as a first-guess period. We can see from the top panel in \autoref{fig:diagnostic} that further error is being introduced into the period estimate by this step, as indicated by the lack of RRc pulsators around the identity line. Even with reasonable first-guess periods, the MCMC template fitting still struggles to ``lock on" to a fit. This has been tested with both the Lomb-Scargle aligned periods from above and the periods output by the HST MCMC fitting. The way the template-fitting code is constructed, the only parameter shared between the individual filter fits is the period. Because the baseline of the F150W filter is less than one full cycle of all of our sources, it struggles to fit even the shortest-period RRc stars. Again, the short baseline of observations combined with the filter switch limits our reasonable parameter space (and introduces possible aliasing) such that the template-fitting cannot find a sufficient absolute minimum. 

In the case where a robust period can be otherwise identified, template fitting can be still performed by fixing the period from the previous step (removing it from the parameter space) and yields robust fits (see \autoref{sec:jwstpwz}).

\subsection{Possible Alternative Uses and Guidance}
While insufficent time baseline observations lead to difficulties following standard methodology, we can establish the following alternative guidance for searching for RRL in the JWST archive, when the method presented in this manuscript is not applicable:
\begin{enumerate}
    \item The Lomb-Scargle, Lafler-Kinman, and combined Saha-Vivas algorithms all reasonably identify proper periods when the time baselines of the observations are sufficent (e.g., the RRc stars in these observations).
    
    \item When the different filters are stitched in a single light curve, the Lomb-Scargle algorithm is roughly able to find the periods for the RRab stars, albeit with greater scatter than would be preferable. A well-calibrated ab-type PWZ could potentially make use of these, along with simple average or midrange magnitudes (see below).
    
    \item Using the co-observed stars, we can find alternative ways to assess apparent brightness which reasonably correlate to the central magnitudes output by the MCMC fit. \autoref{fig:mags} shows the comparison between the period-locked JWST MCMC output and the average and midrange magnitudes for our co-observed sources. Midranges are inspected along with averages, as they are expected to be less impacted by asymmetries in RRab light curves. \autoref{tab:deviations} shows that the filter-wise scatter about the identity function (calculated via the residual standard error with respect to $x=y$) is reduced in the midrange estimate, as compared to the simple average, for both filters. We conclude that—on a population level—midranges can work as a reasonable estimate for RR Lyrae central magnitudes, even in relatively short-duration data, where coverage of both light curve extrema is not guaranteed.

    \begin{figure}
    \centering
    \includegraphics[width=0.9\linewidth]{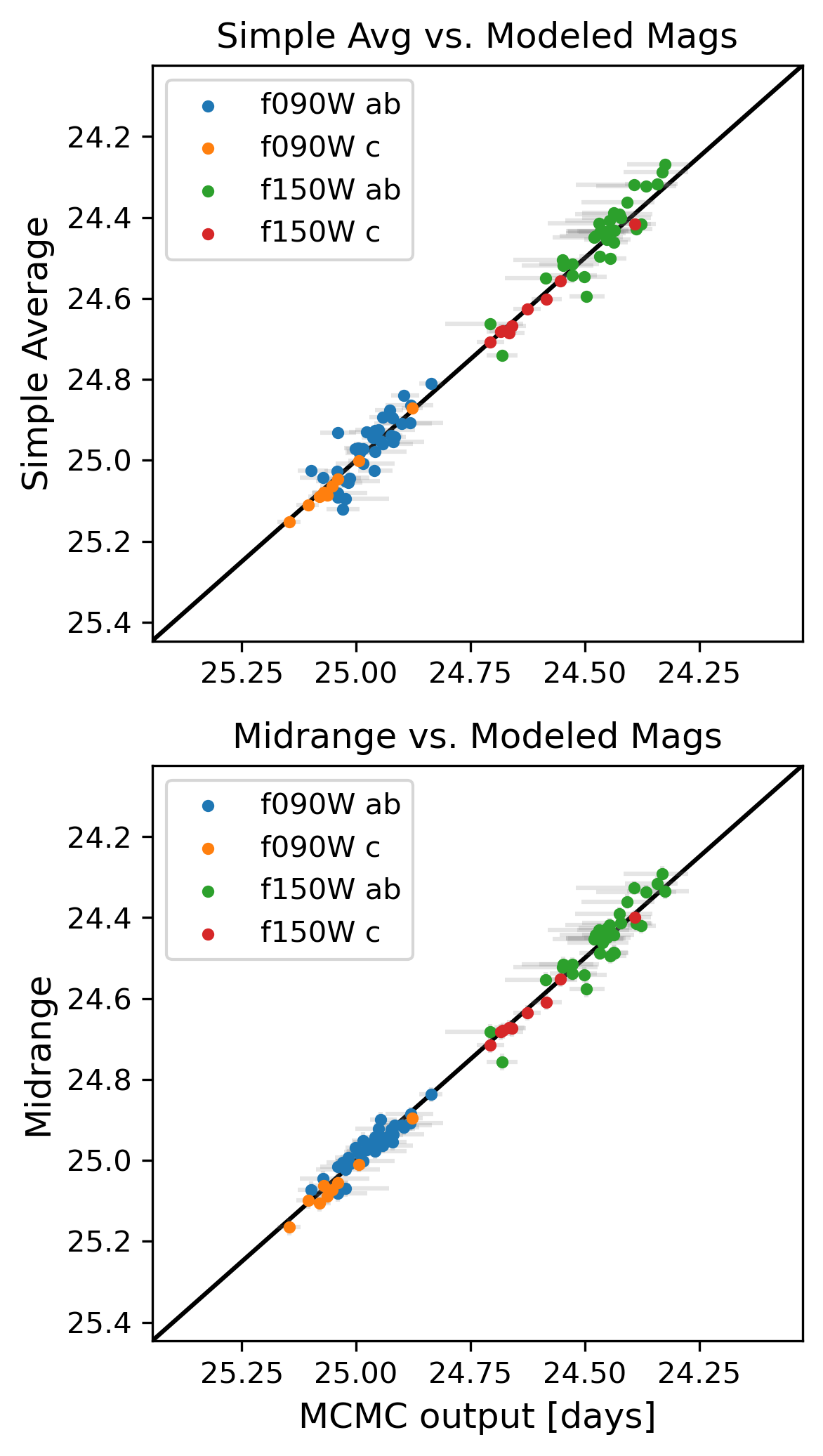}
    \caption{The light curve average and midrange magnitudes versus the mean magnitudes output by the MCMC (when locked on the HST-found periods) for our 44 overlapping stars. The expected value models are indicated by the identity in black.}
    \label{fig:mags}
    \end{figure}
    
    \item All of this has been motivated by trying to use the PWZ method to determine distances. In principle, one could instead compare the observed magnitude of the F090W horizontal branch (calculated as the median of the individual stars' midranges) as a standard candle distance indicator without the need for PWZ fitting.  The uncertainty in this method is impacted by the ``tilting'' of the horizontal branch in the CMD. This tilting is represented by the standard deviation from the filter mean. The filter-wise standard deviations can be found in \autoref{tab:deviations}. 
    We suggest making use of the F090W over the F150W value because the horizontal branch is less sloped in this filter (reflected in the standard deviations in \autoref{tab:deviations}). In the 44 co-observed sources in WLM, we find observed mean magnitude of the F090W horizontal branch (based on the MCMC output) to be $24.99\pm0.070$.
\end{enumerate}

\begin{table}
    \centering
    \caption{Scatter About the Identity}
    \begin{tabular}{lccc}
        \hline\hline
         & F090W & F150W & Total \\
        \hline
        Simple Avg. RSE & 0.039 & 0.038 &0.038\\
        Midrange RSE & 0.023 & 0.033 &0.028\\
        \hline
        Simple Avg. STD & 0.082 & 0.11 &---\\
        Midrange STD & 0.071 & 0.12 &---\\
    \end{tabular}
    \tablecomments{The residual standard errors from the identity function (RSE) and standard deviations about the mean (STD) for the 44 co-observed sources, broken down by filter. In general, the midrange scatter is reduced compared to that of the simple average. As expected, the standard deviation from the mean is less in the F090W filter.}
    \label{tab:deviations}
\end{table}

For the time being, we can still obtain a JWST-HST PWZ distance calibration using the stars in the overlap region. To do this, we re-run the JWST MCMC template fitting, assuming the HST MCMC periods as the true period of the star. The JWST characteristic parameters (RA and Dec coordinates and template-fitting outputs) for the 44 overlap-region RR Lyrae can be found in \autoref{sec:tables}, \autoref{tab:JWSTfinalfits}.

\section{Distance Determination with HST}\label{sec:hstdistance}

To establish an estimated distance to WLM, we fit the RRab data to the period-Wesenheit-metallicity (PWZ) function \citep[as in][]{Savino2022}. The general form of the PWZ is
\begin{equation}\label{eq:pwz_general}
    W_{\rm HST} = \mu + a + b\ \text{log(P)} + c\ \textrm{[Fe/H]}
\end{equation}
 Where W is the Wesenheit magnitude \citep{Madore1982}; $\mu$ is the distance modulus; P is the pulsation period in days; [Fe/H] is the metallicity; and $a$, $b$, and $c$ are calibrated—empirically or theoretically—for a given set of filters, $X,Y$. \autoref{tab:hstconstants} shows the existing calibration for the HST F475W and F814W filters.

Wesenheit magnitudes are defined to be reddening-free:
\begin{equation}\label{eq:W_obs}
    W (X, X-Y) = X - R(X - Y)
\end{equation}
\begin{equation}
    R = A_X / E(X-Y)
\end{equation}
where $R$ is the total-to-selective dust absorption ratio, and X, Y are the mean magnitudes for a given star in the associated filters. For HST, $R_{\rm HST} = 0.960$ as in \cite{Savino2022}. For notation purposes and to avoid ambiguity, we will use subscripts to refer to the calibration and fit values for HST and JWST. Note that a \autoref{eq:W_obs} magnitude can, in principle, be defined for any given combination of filters, given an appropriate extinction law for $R$.

For the HST data fit, we use the standard form of the PWZ from \autoref{eq:pwz_general}, with an additional empirical correction offset, described in \cite{Savino2022}.
\begin{equation}\label{eq:deltamu}
    \Delta\mu = 0.23\text{log(P)} - 0.02\textrm{[Fe/H]} - 0.087(V-I)
\end{equation}

This offset has been derived to match the difference between the empirical PWZ calibration based on Johnson V, I magnitudes and Gaia distances \citep{Nagarajan2022}, and the theoretical PWZ calibration from \citet{Marconi2015}

\begin{table}
    \centering
    \caption{HST F475W and F814W PWZ Constants}
    \begin{tabular}{cc}
        \hline\hline
        Constant & Value \\
        \hline
        $a_{\rm HST}$ & $-0.990\pm0.007$\\
        $b_{\rm HST}$ & $-2.394\pm0.025$\\
        $c_{\rm HST}$ & $0.129\pm0.004$\\
        $R_{\rm HST}$ & $0.960$\\
        \hline
    \end{tabular}
    \tablecomments{Values from \cite{Savino2022}, adapted from \cite{Marconi2015}.}
    \label{tab:hstconstants}
\end{table}

\subsection{PWZ MCMC Fitting}

We again use \texttt{emcee} to conduct an MCMC fit of the PWZ function to our data, following the work of \cite{Savino2022}. For this, the autocorrelation cutoff is defined as in \autoref{sec:HSTmcmctempfit}. 

We adopt a standard Gaussian Mixture Model (GMM) in order to avoid making hard cuts (such as sigma clipping) on outlier data. For each source in the sample, we use our template-fitting-derived mean magnitudes and periods to calculate individual distance moduli, $\mu_k$, using Equations \ref{eq:pwz_general}, \ref{eq:W_obs}, and \ref{eq:deltamu}. We assume the sample of $\mu_k$ are pulled from a normal distribution, $\mathcal{N}(\mu, \sigma^\mu_k)$, where the median $\mu$ is the true distance modulus of WLM, and the standard deviation $\sigma^\mu_k$ is obtained, for each star, through propagation of the measurement uncertainties.

Various sources—such as aliasing and mis-classification—may introduce contamination into our sample. The GMM formalism allows us to account for the effect of outlier stars by modeling them as being members of a second ``false" contaminant population. The contaminant population is drawn from a normal distribution, $\mathcal{N}(\mu_f, \sigma_f)$, where both $\mu_f$ and $\sigma_f$ serve as free nuisance parameters in our model. Instead of enforcing a binary decision as to which population a given star belongs to, we employ a sigmoid-like probability function:
\begin{equation}\label{eq:prob}
    Q_k = \frac{1}{1+\text{exp}(-s(R_k-2))},
\end{equation}
where $R_k = |\mu_k-\bar{\mu}|/\sigma^\mu_k$ is the normalized absolute deviation from the average of the $\mu_k$ distribution. This significantly reduces the constraining power of outlier sources. With this, we can calculate the likelihood of a given parameter set:

\begin{equation}
    \begin{split}
        &p(\mu_k|\mu, \textrm{[Fe/H]}, \mu_f, \sigma_f, s) \\
        &= \frac{(1 - Q_k)}{\sigma^\mu_k (2\pi)^2}\text{exp}(-(\mu_k-\mu)^2 / 2(\sigma^\mu_k)^2 \\ 
        &+ \frac{Q_k}{\sigma_f (2\pi)^2}\text{exp}(-(\mu_k-\mu_f)^2 / 2(\sigma_f)^2
    \end{split}
\end{equation}

The prior distributions of our model parameters can be found in \autoref{tab:hstpriors} \citep[largely adopted from][]{Savino2022}. In general, we assume broad, uninformative, uniform priors. The one exception is for [Fe/H], as outlined in \autoref{sec:metallicities}, wherein we adopt a physically-founded prior.

\begin{table*}
    \centering
    \caption{Prior Distributions for the HST GMM\label{tab:hstpriors}}
    
    \begin{tabular}{ccc}
        \hline\hline
        Param. & Prior & Description \\
        \hline
        $\mu$ & $\mathcal{U}(23,26)$ & Distance modulus of the galaxy \\
        {[}Fe/H{]} & $\mathcal{U}(-2.25,-1.25)$ & Metallicity of the RRL population \\
        $\mu_f$ & $\mathcal{U}(20,30)$ & Dist. modulus of the contaminants\\
        $\sigma_f$ & $\mathcal{U}(0,10)$ & Scatter of the contaminants\\
        s & $\mathcal{U}(1,10)$ & Steepness of the Q sigmoid \\
        \hline
    \end{tabular}
\end{table*}

Using this model, we are able to obtain a value for the distance modulus to WLM for use in our JWST calibration.

\subsection{Metallicity Treatment}\label{sec:metallicities}

Because of the variability in the stellar envelopes and atmospheres of RR Lyrae, spectroscopic determination of metallicities require especially short exposure times, in order to prevent velocity smearing \citep[e.g.,][]{Gilligan2021}. This makes obtaining high resolution spectra especially difficult for stellar populations outside the Milky Way, which are necessarily fainter and therefore typically require longer exposures to reach an acceptable signal-to-noise ratio. In addition, corrections for scatter on metallicity estimates are prone to a number of systematic problems related to determining the associated time-dependent surface gravities \citep{Kovacs2023}.  Recent studies of extragalactic RR Lyrae employ various methods of accounting for the [Fe/H] metallicity in their sources, which often lack sufficient spectroscopic observation. 

Over the past several decades, many studies have sought to create and refine simple functions relating the metallicities of RR Lyrae and the parameters of the Fourier sine (or cosine) decompositions of their light curves \citep[e.g.,][]{Simon1982, Simon1988,Clement1992, Kovacs1995, Jurcsik1996, Morgan2003, Morgan2007, Kovacs2023, Li2023, Muraveva2025}. These include low-order polynomial relationships. In particular, a linear relationship with both the light curve period and the $\phi_{31}$ combined Fourier parameter has been shown to be reasonably robust. These studies have shown to be most effective for evenly- and well-sampled light curves \citep[like those in the OGLE database, e.g.,][]{Soszynski2016}. 

In order to test this relationship as a possible method for adopting a more specific metallicity for our purposes, we create simulated light curves and attempt a Fourier fit to determine $\phi_{31}$ and the associated metallicity. The simulated data are created with the templates used for the MCMC fitting, scaled and repeated to reasonable periods, amplitudes, observing baselines and observing cadences, with random Gaussian noise added similarly in scale to that identified in our photometry, so as to best mimic the real data.

We find that, for a relatively sparsely and unevenly sampled dataset like ours, the first- and third-order phase terms necessary to calculate $\phi_{31}$ are sensitive to noise at a level similar to that in our photometry, as well as errors in the period determination. We find that the spread of metallicity values obtained for various realizations of this noise is greater than our existing ignorance of the population metallicity from SFHs and stellar evolution models, therefore offering no improvement on the treatment \cite{Savino2022} uses for metallicity. As such, we decline to use this treatment at present. While beyond the scope of this work, it may be beneficial to assess a template-fitting Fourier decomposition, as described in \cite{Kovacs2007}, as an alternative in the near future.  

\cite{Savino2022} instead treats [Fe/H] as a free parameter in their PWZ-fitting scheme, assuming a singular population metallicity, instead of looking to solve for the individual metallicities of each source. They use a broad top-hat prior to effectively capture the ignorance of the metallicity in their final model uncertainties. The assumption that the entire population of RR Lyrae have the same metallicity is a first-order approximation, but does not meaningfully impact our results. A spread of metallicity values would manifest as an increased spread in the Magnitude-Period plot, while the best-fitting distance would remain unchanged. In addition, the single-metallicity assumption is as physically founded or better than a number of assumptions that would need to be employed to circumvent the issue.

\cite{Sarajedini2023} makes use of a simple linear trend relating only the AB-type RR Lyrae period with the metallicity: 
\begin{equation*}
    \textrm{[Fe/H]}_{ZW} = -7.82\, \log(P_{ab}) - 3.43
\end{equation*}
with an additional conversion from the \cite{Zinn1984} metallicity scale to the more recently calibrated UVES scale of \cite{Carretta2009}. While the metallicities are calculated on a star-by-star level, this relation is best used to find a mean population metallicity.

For our purposes, we make use of the top-hat prior method from \cite{Savino2022}, where we continue to use the same top-hat width to reflect a similarly conservative estimate of our uncertainties. However, the median has been shifted to $-1.75$, to more closely reflect our knowledge from the star formation history and age-metallicity relationship derived by \citet{McQuinn2024}, as well as the spectroscopically-derived metallicity distribution function presented in \citet{Leaman2009}. 

The summary statistics of the \cite{Sarajedini2023}-like metallicity distribution yield a similar width to the \cite{Savino2022} top-hat. Our data have little informative power over the metallicity prior, and tests with a Gaussian prior show an impact on the final posterior distribution shape, making the uniform top-hat preferable. Further discussion on possible future approaches to addressing the metallicity can be found in \autoref{sec:futurework}.

\section{JWST PWZ Fit and Calibration to HST}\label{sec:jwstpwz}

From here, we can take the distance modulus from our HST fit, and use it as an input to fit our PWZ relation to the JWST data. The fitting process is similar to \autoref{sec:hstdistance}, but in reverse. In order to calibrate the PWZ in the F090W and F150W bands to our \textit{Gaia}-consistent HST relation, we would like to find the constants $a$, $b$, and $c$ from \autoref{eq:pwz_general}. However, without having proper measurements of the stellar metallicities, $a$ and $c$ in the general form of the PWZ are degenerate; we cannot leave [Fe/H] as a free parameter as we have done previously. Therefore, for the JWST data, we re-write our PWZ to combine the degenerate terms:
\begin{equation}\label{eq:pwzalphab}
    W_{JWST} = \mu + \alpha + b_{JWST}\ \text{log(P)}.
\end{equation}
Where $\alpha = a_{JWST} + c_{JWST}\textrm{[Fe/H]}$ could be used to determine $a_{JWST}$ and $c_{JWST}$ in principle. 

Here, $\alpha$ and $b_{JWST}$ will be our test parameters. In order to circumvent the need to implement a 2-dimensional GMM, we instead can take the difference between our two values of W: the one derived from photometric data ($W_{obs}$, \autoref{eq:W_obs}), and the one from the PWZ in \autoref{eq:pwzalphab} ($W_{mod}$). The difference, $\Delta W_k$, is assumed to be sampled from the normal distribution $\mathcal{N}(0, \sigma^{\Delta W}_k)$ where, as before, $\sigma^{\Delta W}_k$ is calculated for each star by propagation of uncertainties. 

Within this framework, the construction of the GMM and associated sigmoid function are analogous to those used for the HST fit. However,  now the parameters and values based on $\mu$ are replaced with those based on $\Delta W$, where again, $\Delta W$ is calculated from the test parameters $\alpha$ and $b_{JWST}$. 

The parameter priors for the JWST PWZ fit can be found in \autoref{tab:jwstpriors}. Again, we establish broadly uninformative, uniform priors.

\begin{table*}
    \centering
    \caption{Prior Distributions for the JWST GMM\label{tab:jwstpriors}}
    \begin{tabular}{ccc}
        \hline
        \hline
        Param. & Prior & Description \\
        \hline
        $\alpha_{JWST}$ & $\mathcal{U}(-10,10)$ & Zero-point offset \\
        $b_{JWST}$ & $\mathcal{U}(-10,10)$ & Line slope \\
        $\Delta W_f$ & $\mathcal{U}(-10,10)$ & Difference in W for the contaminants \\
        $\sigma_f$ & $\mathcal{U}(0,10)$ & Contaminant pop. scatter \\
        s & $\mathcal{U}(1,10)$ & Steepness of the Q sigmoid \\
        \hline
    \end{tabular}
    
\end{table*}

\section{Identified RR Lyrae Populations}\label{sec:rrlpops}

\begin{figure}
    \centering
    \includegraphics[width=\linewidth]{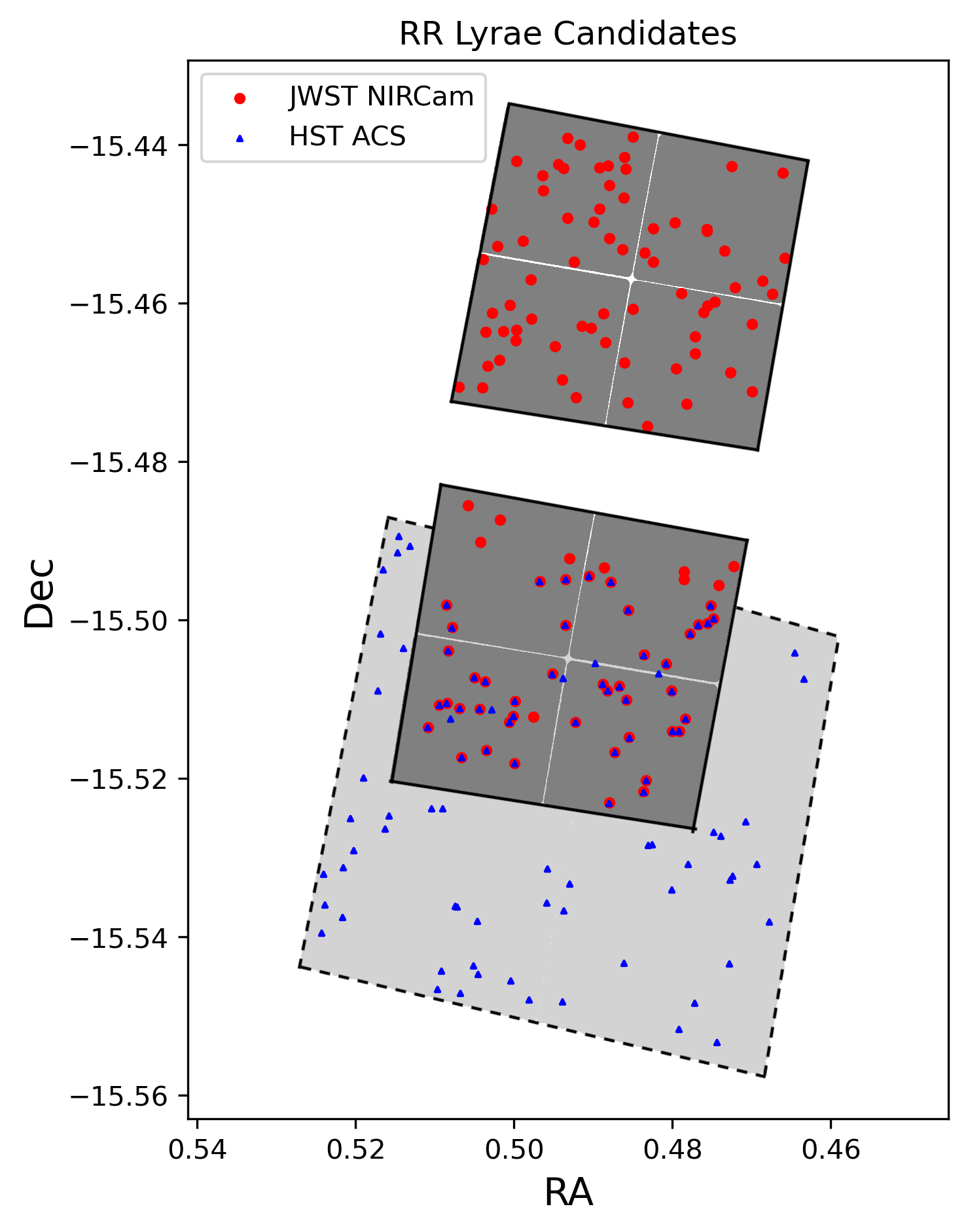}
    \caption{A sky-view plot of our final dataset. The footprints of the HST ACS and JWST NIRCam observations are shown in light and dark gray, respectively, while the identified RR Lyrae are in blue (HST) and red (JWST). Note the three HST sources in the general overlap region which are not seen in \autoref{fig:chi2check}. These sources lie between the NIRCam chips, and were therefore excluded from the cross-matched source analysis. In addition, two of the HST-only sources that are shown in \autoref{fig:chi2check} are not pictured here. The first was the source removed for quality cuts in the data, as described in \autoref{sec:falsepositives}. The second was a relevant false-positive for that analysis, but was ultimately one of the two removed from the PWZ-fitting for high template-fitting uncertainty.}
    \label{fig:finalregions}
\end{figure}

Our final catalogs of RR Lyrae are plotted in \autoref{fig:finalregions}. The HST set contains 101 sources, of which 85 are best fit as RRab, and 16 are RRc. We note that this methodology does not identify double-mode pulsator (RRd) sources. Previous research \citep[e.g.,][]{Alcock2000, Soszynski2016, Soszynski2016b, Clementini2023, Nemec2024, Zhang2025} has determined that RRd sources typically have primary periods associated with first-overtone (RRc) pulsation, and the amplitudes of the pulsations associated with the secondary periods are diminished compared to those of the first. For our purposes, with sparsely-sampled light curves, this means that any RRd sources which may exist in this data set are \textit{most likely} to be best-fit as RRc sources, assuming they are not discarded due to poor light-curve fitting \citep{Savino2022}.

These populations are similar to the populations identified in \cite{Sarajedini2023}, who finds 76 RRab- and 14 RRc-type. \cite{Sarajedini2023} also reports a mean ab period of $\langle P_{ab}\rangle = 0.609\pm0.058$. This study similarly finds a value of $\langle P_{ab}\rangle = 0.61\pm0.03$, well within the margin of error. This mean period puts WLM squarely in the middle of the Oosterhoff I and II classifications \citep[$\langle P_{ab}\rangle \approx 0.65d$ and $\langle P_{ab}\rangle \approx 0.55d$, respectively;][]{Oosterhoff1939, Oosterhoff1944, Catelan2009}. It is expected that WLM would fall, like other dwarf galaxies, in this so-called ``Oosterhof gap" \citep{Catelan2009, Monelli2022}.

\begin{figure}
    \centering
    \includegraphics[width=\linewidth]{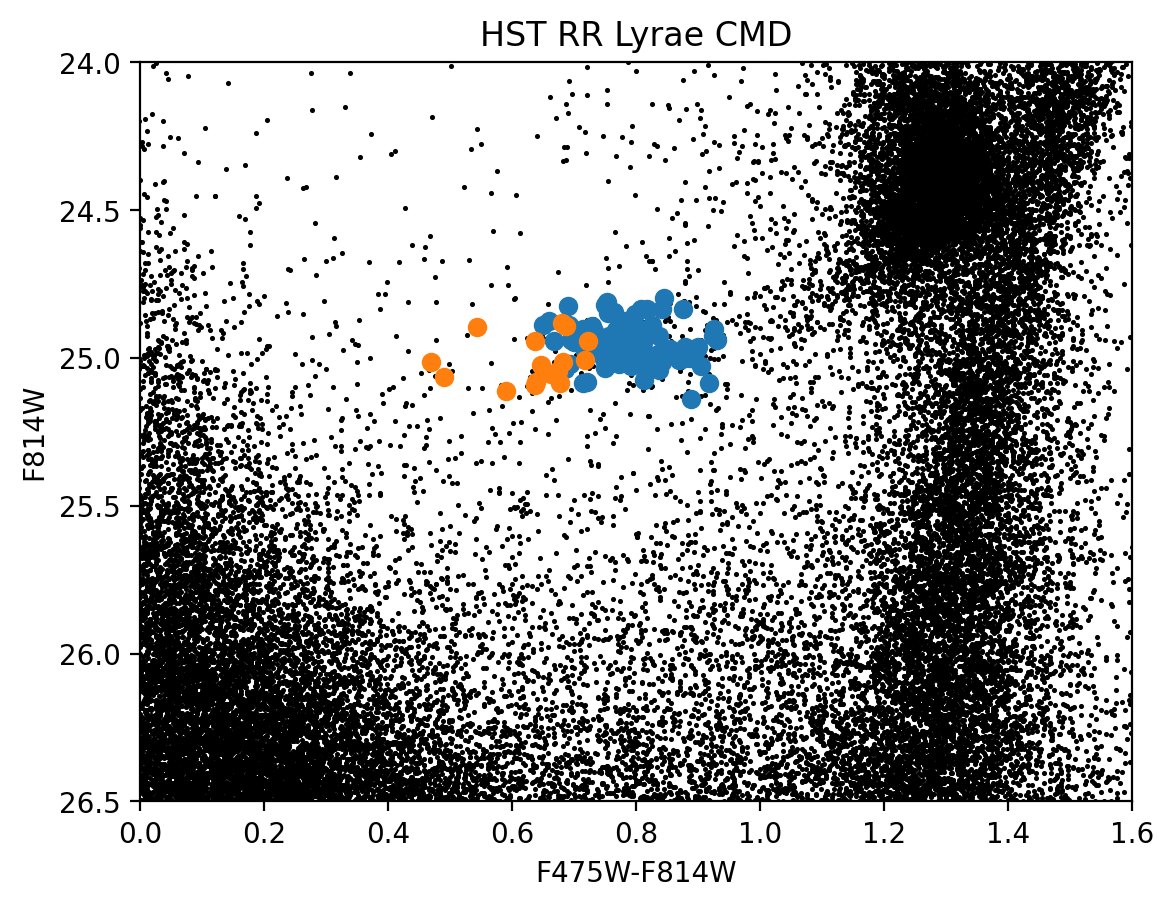}
    \caption{The full HST catalog CMD (black), zoomed in on the horizontal branch, overlaid with the final 101 RR Lyrae stars. RRab stars are shown in blue, while RRc stars are shown in orange.}
    \label{fig:hstoutputcmd}
\end{figure}

Our present set of JWST-observed expected RR Lyrae contains 125 sources, the specific classifications of these are undetermined, given the uncertainty presented by the template-fitting step.
Fortunately, there are 44 co-observed RR Lyrae candidates in both data sets. Of these, 35 are identified in the HST analysis as RRab-type, and 9 as RRc-type. We make use of the 35 ab stars for our PWZ calibration.

\section{Discussion of PWZ Fits}\label{sec:discussion}

\subsection{HST Adopted Distance}

\begin{figure}
    \centering
    \includegraphics[width=\linewidth]{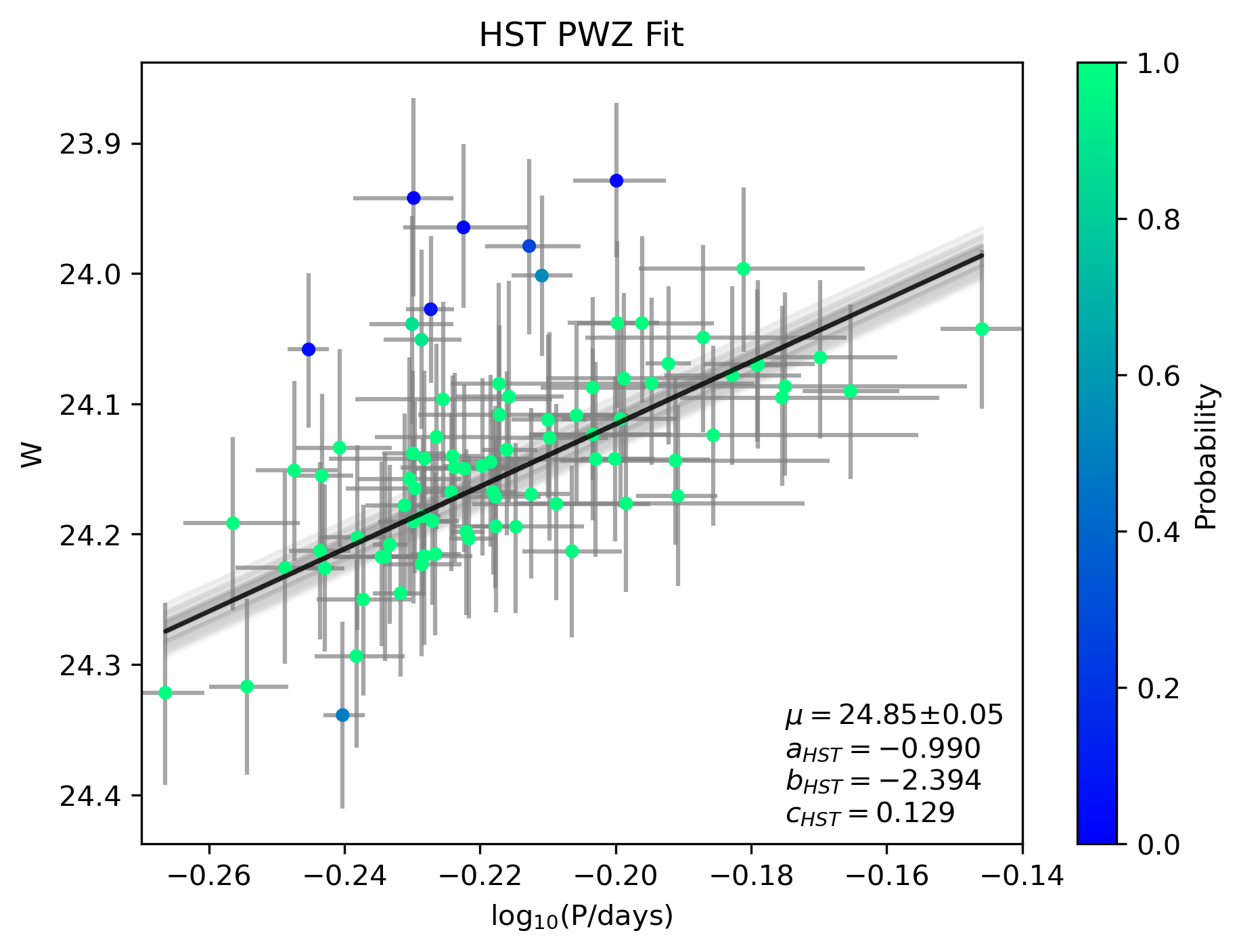}
    \caption{The PWZ fit for the HST data, which is based on 85 RRab stars in the HST/ACS sample. The median fit line is presented in black, while a random subset of other MCMC samples are in gray. Sigmoid-like probability of belonging to the population generated by the median-fit line (\autoref{eq:prob}) is indicated by color. Note the handful of brighter, low-probability possible contaminants above the bulk of the sample.}
    \label{fig:hstpwz}
\end{figure}

The PWZ fit to the HST data is shown in \autoref{fig:hstpwz}. The median fit line is presented in black, while a random subset of other MCMC samples are in gray. The sigmoid-like probability of belonging to the population generated by the median-fit line (\autoref{eq:prob}) is indicated by color. Our sample is similar to the ``populated, noisy sample'' fit presented in \cite{Savino2022}. We see a handful of brighter, low-probability possible contaminants above the bulk of the sample, while the majority of our sources are well distributed about the best-fit line, with similar slope. 

From this, the median distance modulus obtained is
\begin{equation*}
\mu = 24.85 \pm 0.05,
\end{equation*}
where the upper and lower bounds represent the 16th and 84th percentiles in our MCMC posterior. This corresponds to a median physical distance of 0.93 $\pm$ 0.02 Mpc.

\begin{figure}
    \centering
    \includegraphics[width=\linewidth]{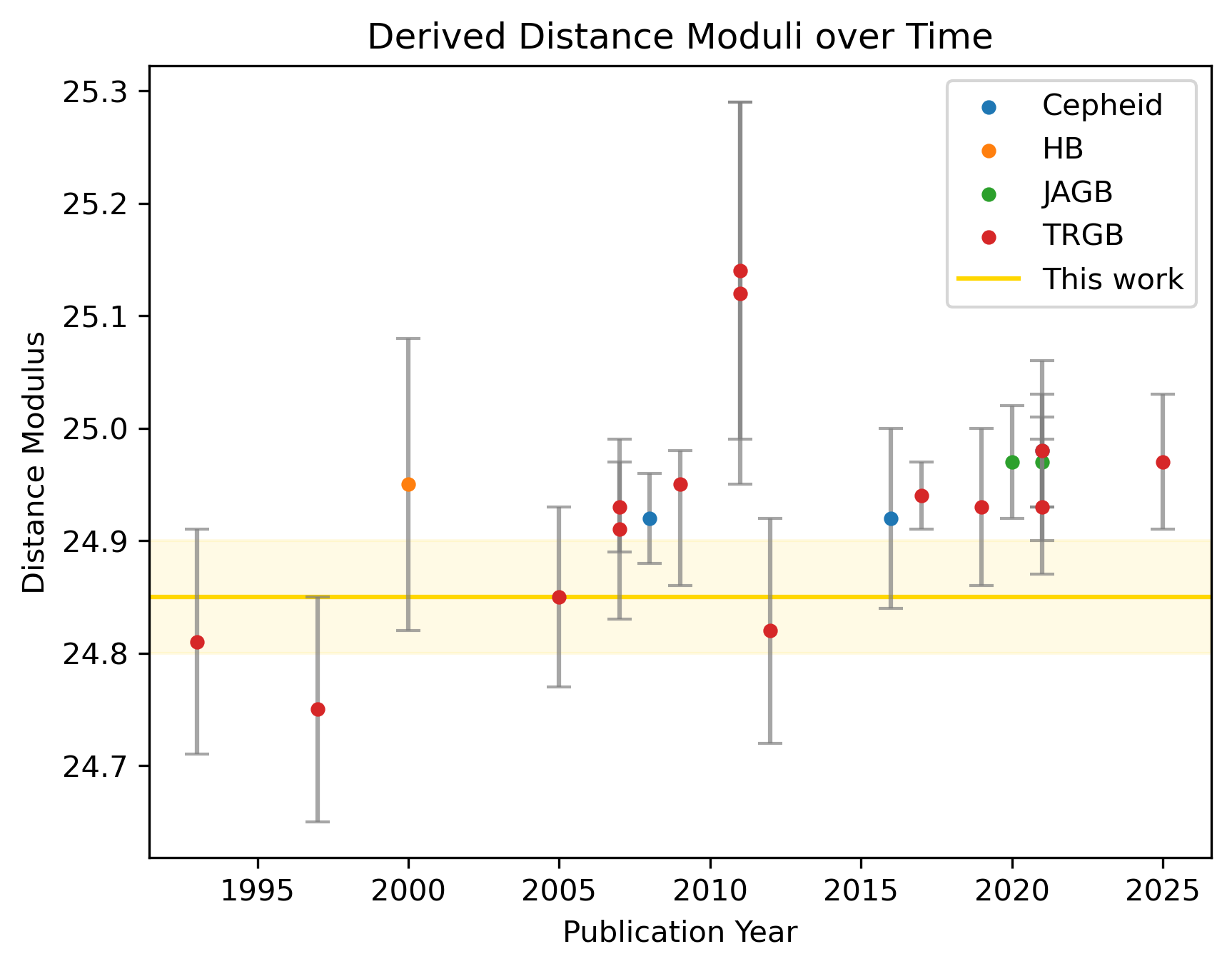}
    \caption{Previous distance modulus results, derived using Cepheid (blue), Horizontal Branch (orange), JAGB (green), and TRGB(red) methods, as compared to the distance modulus obtained in this work (yellow line). 1$\sigma$ uncertainties are indicated by gray error bars (past work) and the yellow shaded region (this work)}
    \label{fig:tension}
\end{figure}

Our final RR Lyrae distance modulus is shown in \autoref{fig:tension}, compared to a number of previous distance results \citep{Lee1993, Minniti1997, McConnachie2005, Rizzi2007, Pietrzynski2007, Gieren2008, Jacobs2009, Gorski2011, McCall2012, Bhardwaj2016A, McQuinn2017, Albers2019, Freedman2020, Lee2021, Yan2025}. In general, our derived distance is somewhat less than previous determinations. For example, \cite{Lee2021} find JAGB, NIR TRGB, F814W TRGB, and Cepheid distances of 
$24.97\pm0.04$, 
$24.98\pm0.08$, 
$24.93\pm0.06$,
and $24.98\pm0.05$, respectively (where the statistical and systematic errors have been added in quadrature).
Additionally, the most recent NIR TRGB distance measurement to WLM was conducted by \citet{Yan2025}. They find a JWST-derived distance modulus of  
$24.977 \pm 0.059$ (where again the statistical and systematic errors have been added in quadrature). These differences represent less than a $2\sigma$ tension, and are therefore not a present cause for great concern.

Again, it is expected that the bulk of the uncertainty in our determination stems from the uncertainty in the metallicity. In order to assess if this is sufficient to account for the present tension, we can reverse our PWZ calculation. Adopting an assumed distance modulus of $24.96$ (the weighted average of the \cite{Lee2021} values), and plugging into \autoref{eq:pwz_general} (with the empirical correction) we can solve for $\textrm{[Fe/H]}$ for each star to roughly estimate the metallicities required to account for the tension. 
Iteratively sigma-clipping the distribution yields a mean expected $\langle \textrm{[Fe/H]}\rangle = -2.49 \pm 0.21$. This is significantly below the estimated mean metallicity obtained by \cite{Sarajedini2023}. It is unlikely, therefore, that the the distance tension with \cite{Lee2021} is due to our uncertainty in metallicity.

It is most likely, then, that the majority the tension is related to differences in calibration. \cite{Lee2021} make use of an absolute distance calibration to the LMC based on detached eclipsing binaries \citep{Pietrzynski2019}, while our determination is consistent with \textit{Gaia}. One can get a sense of the difference incurred by consistency with Gaia from \cite{Savino2022}, who find the addition of the Gaia term yields distances $\sim 3\%$ closer. By comparing the theoretical PWZ from \cite{Marconi2015} with Gaia-consistent distances, they deduced a period-dependent difference which ranged from $-$0.02 to $-$0.10, which, for a fully sampled RR Lyrae population would result in a roughly 0.04 mag closer distance modulus. An offset of this magnitude fully explains the difference between our distance determination and those of \citet{Lee2021}. Similarly, \citet{Yan2025} make use of an F090W TRGB magnitude calibrated via the megamaser host galaxy NGC 4258 \citep{Humphreys2013, Reid2019, Anand2024}.

\subsection{JWST Calibration}
\begin{figure}
    \centering
    \includegraphics[width=\linewidth]{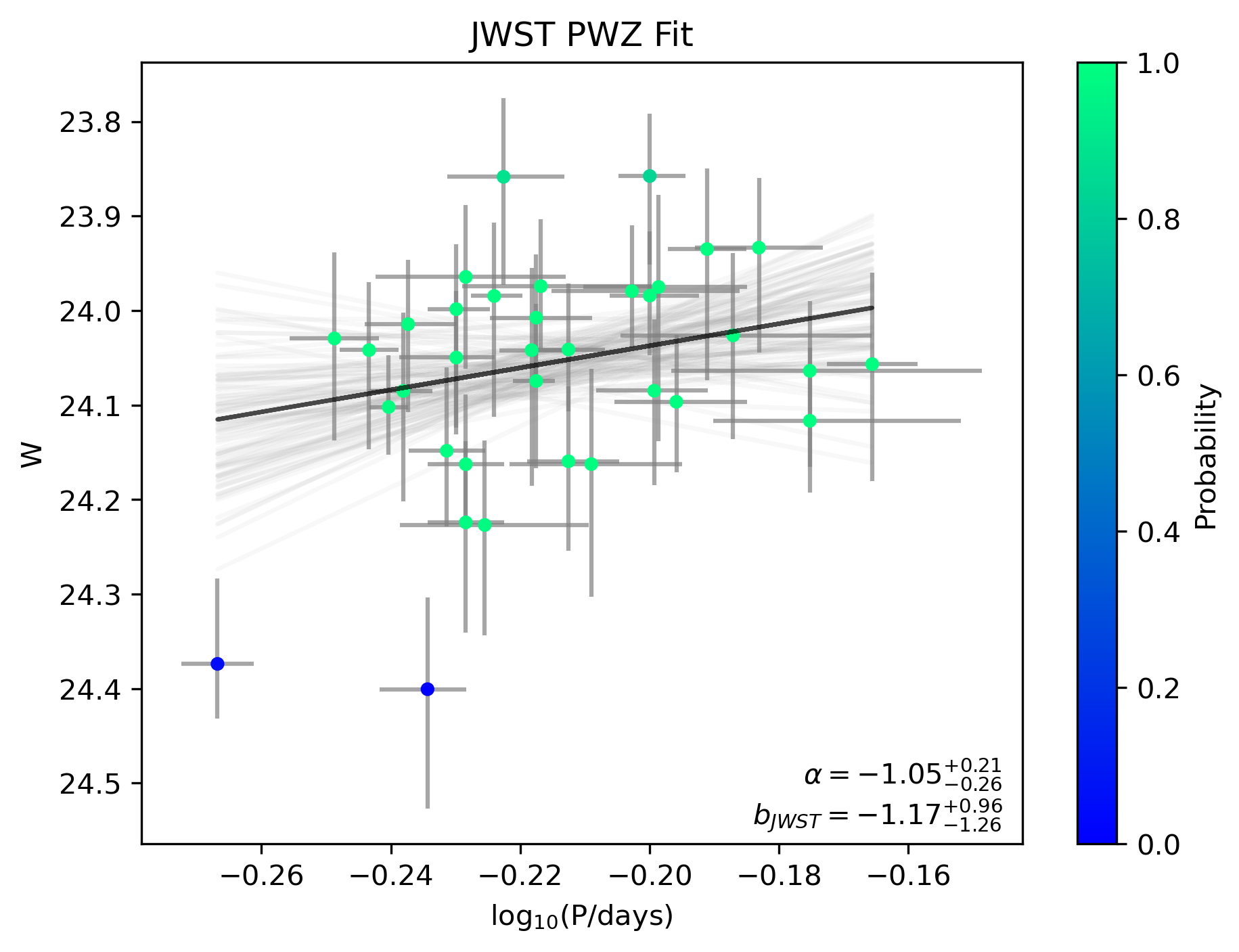}
    \caption{The PWZ fit for the JWST data, which is based on 35 RRab stars identified in both the HST and JWST data}. The median fit line is presented in black, while a random subset of other MCMC samples are in gray. As compared to the HST fit, this random subset is much more spread out, a result of the increased error in the calibration parameters. The sigmoid-like probability of belonging to the population generated by the median-fit line (analogous to \autoref{eq:prob}) is indicated by color.
    \label{fig:jwstpwz}
\end{figure}

The JWST fit is shown in \autoref{fig:jwstpwz}, where again, the median fit line is presented in black, a random subset of other MCMC samples are in gray, and the sigmoid-like probability of belonging to the population generated by the median-fit line is indicated by color. This fit finds calibration values of 
\begin{equation*}
\begin{split}
    \alpha &= -1.05 ^{+0.21} _{-0.26} \\
    b_{JWST} &= -1.17 ^{+0.96} _{-1.26}.
\end{split}
\end{equation*}
These values have significantly higher uncertainty than the analogous calibration in HST (see \autoref{tab:hstconstants}). The increase in uncertainty can be seen as a greater spread in the random selection of MCMC samples in \autoref{fig:jwstpwz}. This is likely a result of our limited sample size and increased scatter in the data impacting our calibration, creating a greater degeneracy between the slope and intercept of the line when trying to identify best-fit values. 

Note that the slope is not well constrained. However, the prior on the slope, $b_{JWST}$, is open to a broad range of positive values, and the MCMC consistently converges on the negative value presented here.

\section{Conclusions}\label{sec:conclsions}

We have conducted an analysis of the RR Lyrae population observed by HST in nearby dwarf galaxy WLM. In these data, we identified variable sources, measured their periodicity, and fit their light curves to existing templates in order to obtain the parameters of variability for each. Using these fits, we calculated the RR Lyrae PWZ distance to WLM.

Additionally, we have laid the groundwork for future studies of RR Lyrae using the JWST NIRCam instrument. We again identified variable sources in WLM, and are able to quantify a general relative success rate of this identification. For the periodicity-measuring and template-fitting steps, we have discovered systematic issues with the JWST data that will be a priority in future work. While the lower variability amplitude in the NIR plays a small role, the main limitation of these observations is the short time baseline.  Here we have investigated a number of alternative solutions for use with existing, short-baseline archival data. We can additionally recommend that, under the NIRCam filter wheel constraints, similar future stellar population studies will need to divide the observations into multiple epochs (as is typical for variable star studies). 
While this would increase the time baseline of observations, without the need for additional integration time, it would, of course result in an additional overhead cost.

Finally, we used the sources identified in both the HST and JWST data to create a first-of-its-kind calibration of the PWZ to JWST filter bands. This calibration can be improved upon with simultaneous fitting to additional populations.

\subsection{Future Work}\label{sec:futurework}

To obtain a more secure calibration of the JWST PWZ, this work needs to be extended to other galaxies. For example, there are suitable HST JWST observations of Leo P \citep{McQuinn2015, McQuinn2024} and the JWST observations have a duration of 36 hours obtained over 4 days, over twice the integration time of our WLM JWST data. 

Secondly, we are interested in testing different prior distributions for our PWZ MCMC fitting. In particular, we could replace our $\alpha$ and $b$ with physically-motivated, weakly informative Gaussian priors. Doing so may reduce the error in our fits. The ground-based $W(H,I-H)$ coefficients (where H and I are similar to our JWST bands) presented in \cite{Marconi2015} could provide the physical motivation.

The bulk of the uncertainty in our distance modulus stems from our ignorance of the individual metallicities of our RR Lyrae. While assuming a bulk group metallicity has proven to be reasonably effective in past studies \citep[for additional discussion, see][]{Savino2022}, recent studies of Milky-Way RR Lyrae \citep{Bhardwaj2023} are strengthening our understanding of the period-metallicity relationship. Such an understanding could allow us, at minimum, to make first-order corrections to the singular metallicity estimate, such as a metallicity gradient treatment. In addition, studies of age/metallicity relationships for extragalactic RR Lyrae could improve our metallicity estimates. OGLE observations of the LMC and SMC, in conjunction with the Fourier analysis of \cite{Kovacs2023}, would be well-suited to this end.

More broadly, we are interested in better assessing the completeness of our recovered sample in JWST, building off of the preliminary analysis from \autoref{sec:probabilities}. To do this, we can use our light curve templates to create a suite of artificial RR Lyrae, pulling from the full expected RR Lyrae period-amplitude parameter space. By running these stars through the analysis outlined here, we can better quantify our recovery rates for different parameter combinations. In addition, we can also test, for a range of observational baselines and periods, the likelihood of observing sections of the RR Lyrae light curves that will register as variable in our identification steps. This would help generalize our results from \autoref{sec:probabilities}. This would allow us assess how long an observational baseline is needed to reasonably observe these sources, a reference with enormous potential utility for future observing practices.

\section*{Acknowledgments}
The authors extend special thanks to Abhijit Saha for his input and consultation on this work. C.S. wishes to thank Vittorio Braga for providing the normalized H-band templates used in this paper.

This work is based on observations with the NASA/ESA/CSA James Webb Space Telescope obtained (program ERS-DD-1334) from the Data Archive at the Space Telescope Science Institute, which is operated by the Association of Universities for Research in Astronomy, Incorporated, under NASA contract NAS5-03127.

This work is also based on observations with the NASA/ESA Hubble Space Telescope (program GO-13768) obtained from the Data Archive at the Space Telescope Science Institute, which is operated by the Association of Universities for Research in Astronomy, Incorporated, under NASA contract NAS5-26555.

Additional support for this work was provided through a grant from the STScI under NASA contract NAS5-03127, program number JWST-AR-03248 (PI: Skillman).

This research has made use of NASA's Astrophysics Data System Bibliographic Services. All of the JWST and HST data used in this paper can be found in MAST: \dataset[10.17909/vd4r-hn91]{http://dx.doi.org/10.17909/vd4r-hn91}.

\facilities{HST (ACS), JWST (NIRCam)}
\software{This research made use of routines and modules from the following software packages: 
\texttt{AstroPy} \citep{Astropy:2013,Astropy:2018,Astropy:2022},
\texttt{DOLPHOT} \citep{Dolphin2000, Dolphin2001, Dolphin2004, Dolphin2016, Weisz2024},
\texttt{emcee} \citep{Foreman-Mackey2013},
\texttt{Matplotlib} \citep{Matplotlib:2007},
\texttt{NumPy} \citep{Numpy:2020},
\texttt{Pandas} \citep{McKinney_2010, McKinney_2011},
\texttt{Psearch} \citep{Saha2017},
and \texttt{SciPy} \citep{Scipy:2020}
}

\appendix
\vspace{-0.8cm}
\section{Analysis of Systematic Zero-Point Offsets in JWST Data}\label{sec:appendix}

It is important, when dealing with time-series data such as these, to consider the potential impact of systematic zero-point offsets introduced by the mode(s) of data collection by the telescope. These have been studied and (when necessary) calibrated for in HST. Here, we outline an initial analysis of such offsets in JWST, and demonstrate that the impact on the results presented here is minimal.

To assess this, we begin with the full catalog of $\sim1.7$ million sources from JWST, as output by \texttt{DOLPHOT}, and cull it with guidance from the documentation\footnote{https://dolphot-jwst.readthedocs.io/en/latest/post-processing/catalogs.html}. We first select based on sharpness, crowding, photometric quality flags, signal-to-noise ratio (SNR), and object type, where the first four are conducted independently for each NIRCam filter. The cutoff values are found in \autoref{tab:zpcutoffs}, and are established to prioritize purity as in \cite{Weisz2024}. It should be noted that while we include the type and quality flag selections, they do not remove any sources that are not otherwise caught.

\begin{table}[h]
    \centering
    \caption{JWST Photometry Selection}
    \begin{tabular}{cc}
     \hline
     \hline
     Selection & Cutoff \\
     \hline
     $\text{SHARP}^2$ & $\le0.01$\\
     CROWD & $\le2.25$\\
     FLAG & $\le0$ \\
     SNR & $> 50$\\
     OBJTYPE& $\le2$\\
     \hline
    \end{tabular}
    \tablecomments{The selection cutoffs for our JWST photometry. Note that SHARP, CROWD, FLAG, and SNR are each applied filter-wise.}
    \label{tab:zpcutoffs}
\end{table}

\begin{figure}
    \centering
    \includegraphics[width=0.45\linewidth]{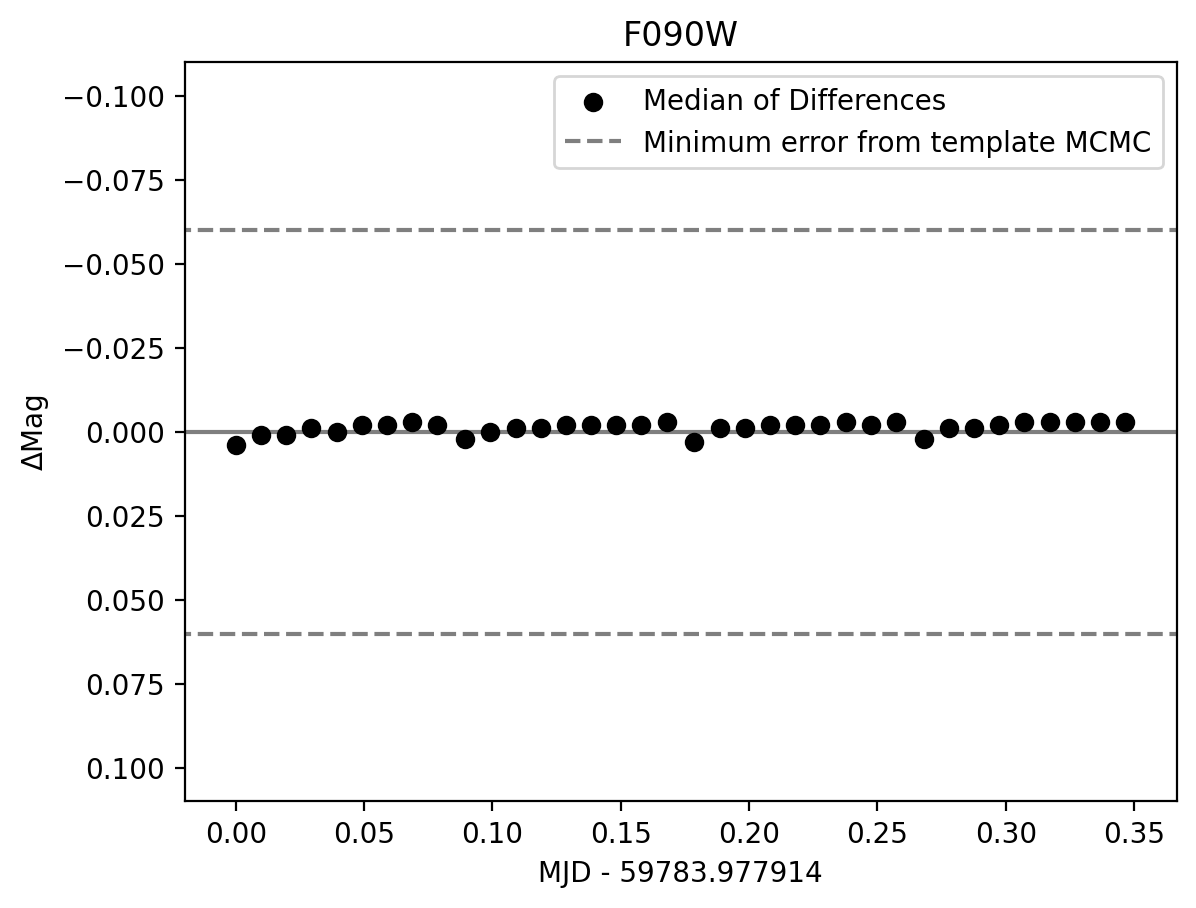}
    \includegraphics[width=0.45\linewidth]{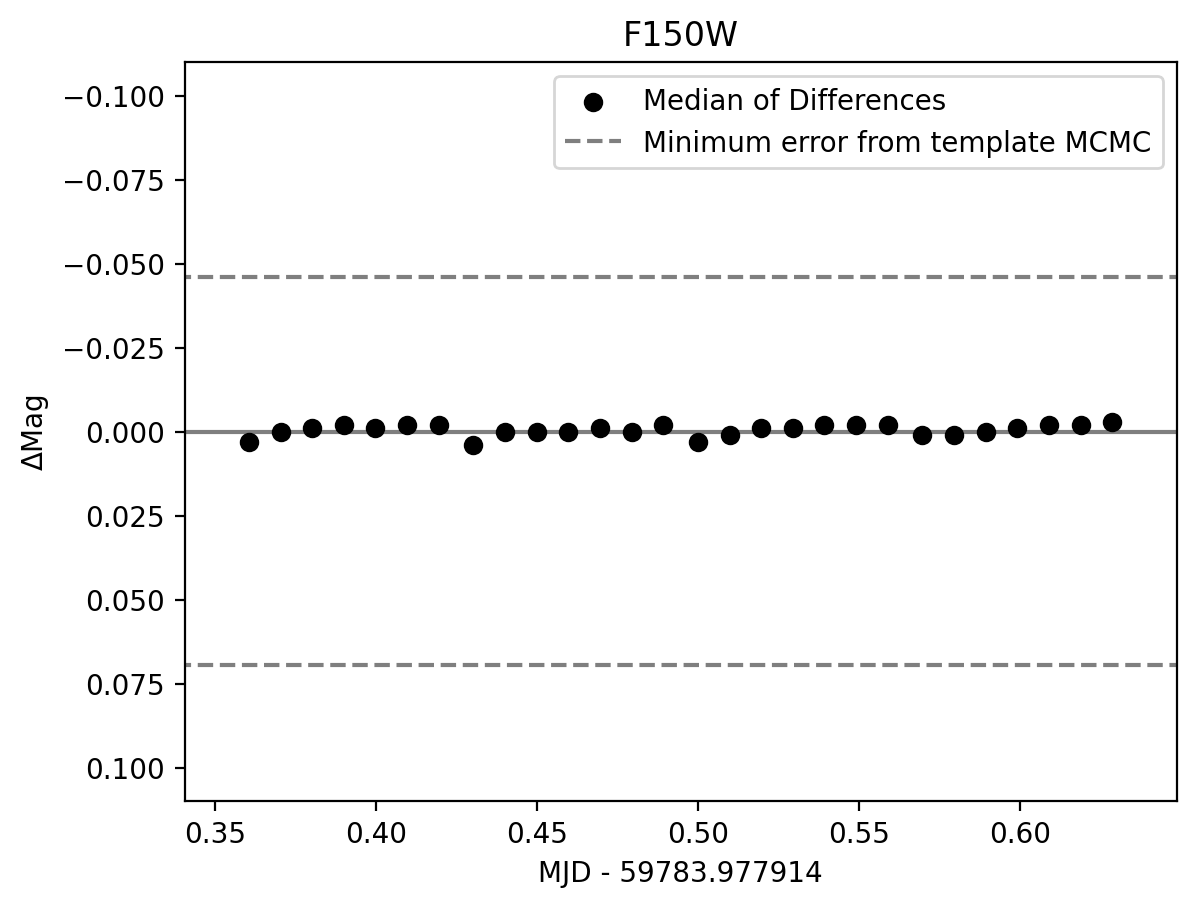}
    \caption{The median difference per epoch vs.\ time for both JWST filter data sets (F090W, left; F150W, right). In both, the zero-point is indicated by a solid line, and the minimum (asymmetric) errors on the amplitude, as output by the template-fitting MCMC, are indicated by dashed lines.}
    \label{fig:JWSTzp}
\end{figure}

Our remaining catalog contains $\sim74$ thousand sources with pristine photometry. We split this catalog between the F090W and F150W filters. We subtract the instrumental VEGAMAG magnitude value from our final photometric catalog from the value at each epoch, so that each source has a set of magnitude differences. Finally, we take the median difference across all the sources in a given epoch and compare the median differences with the amplitude uncertainties output from the template fitting. \autoref{fig:JWSTzp} shows the median differences vs.\ time, as well as the minimum template-fitting errors, for both JWST filters.

The JWST NIRCam photometry appears to be exceptionally stable.  When comparing to the minimum, worst-case-scenario, uncertainties from the template fitting routine, we find that the median difference in zero-point is much smaller. At most, the median systematic zero-point offset is $0.004$mag in either filter. As such, we do not expect the contribution of systematic zero-point offset to the overall error on the template fits to be consequential for our purposes. We note that such an effect may need to be accounted for in future time-series research, if the errors on the scientific products are sufficiently small.

While the magnitude of the JWST systematic zero-point offset is not large, it is of interest to point out that it clearly follows a pattern in time. In both filters, the data appear as distinct ``sets" of consecutive observations. In each set, the first epoch has a generally lower median zero-point, and the offsets increase with time, until the next set begins. The ``reset'' coincides with the dithering of the observations.

\begin{figure}
    \centering
    \includegraphics[width=0.45\linewidth]{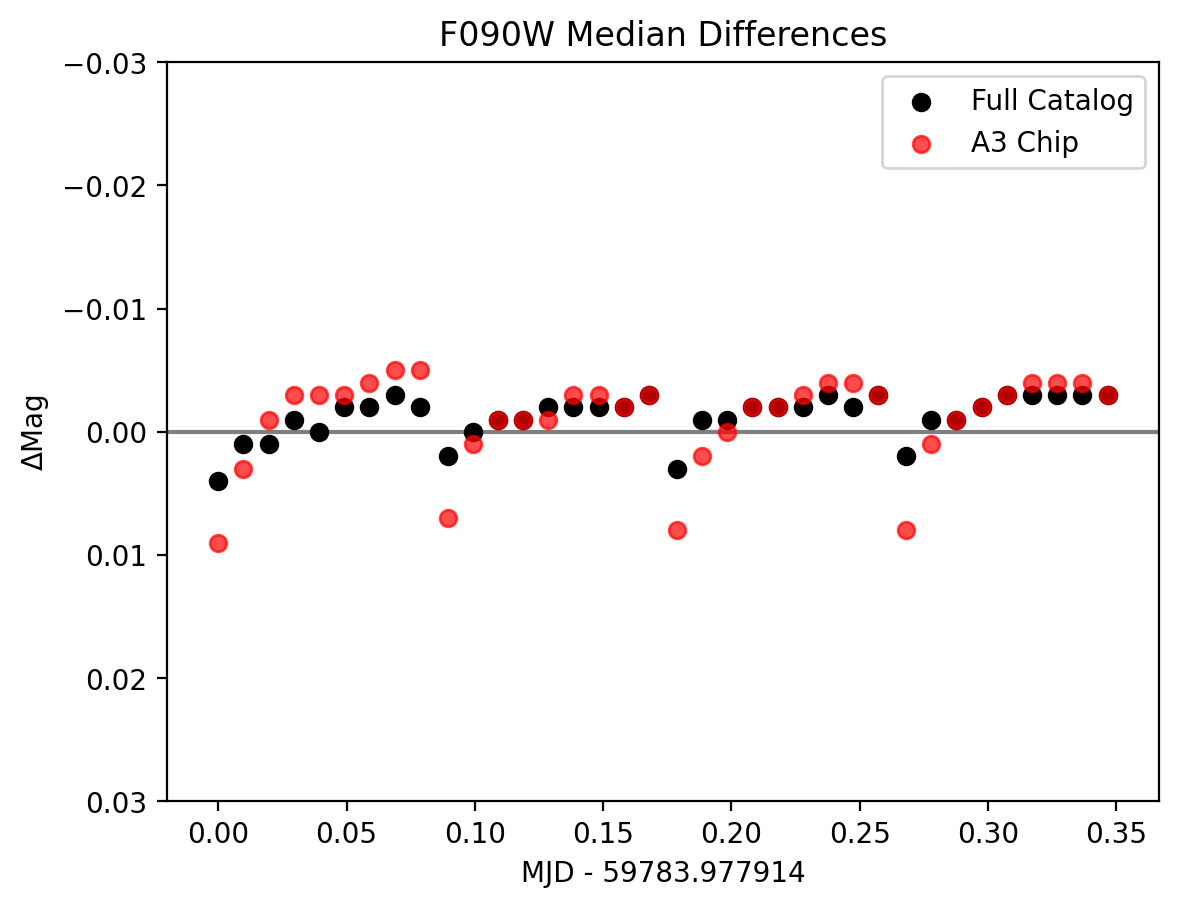}
    \includegraphics[width=0.45\linewidth]{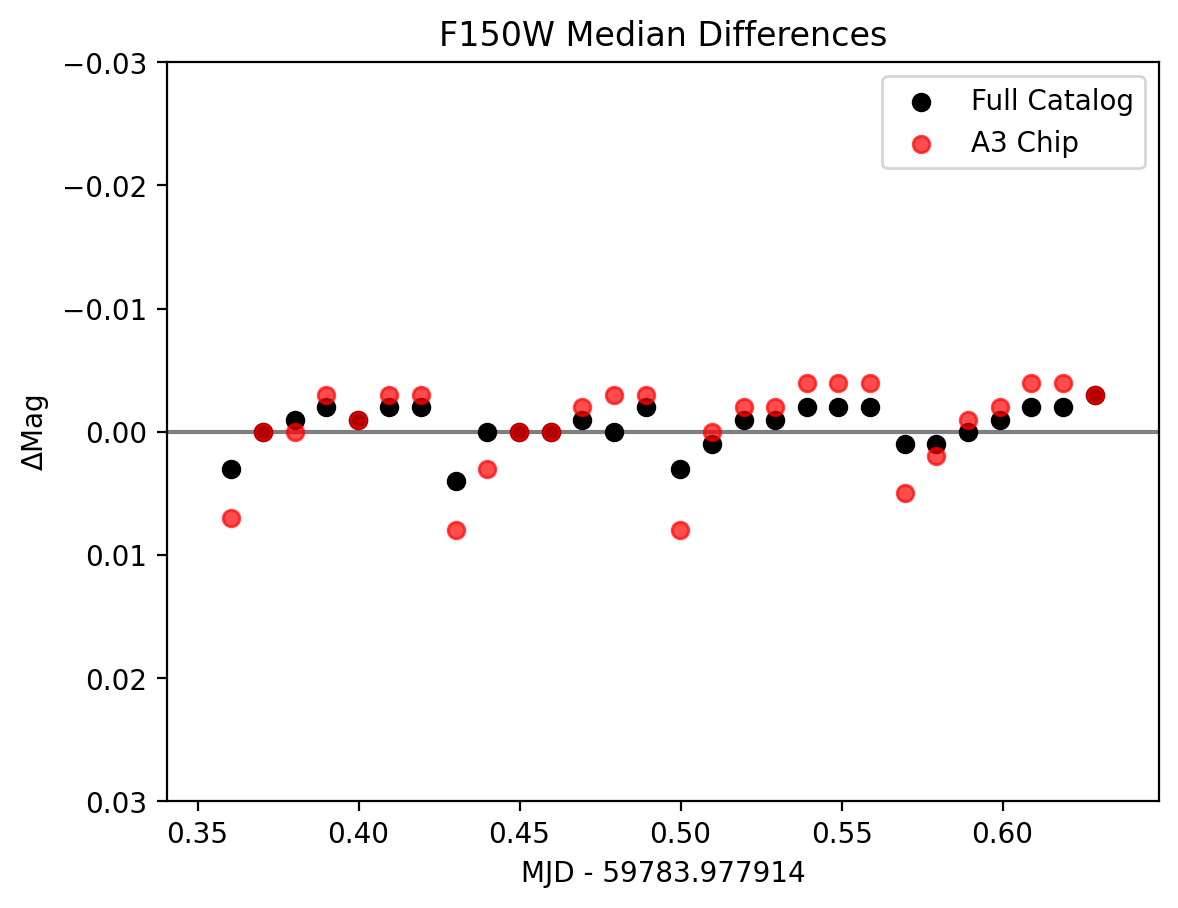}
    \caption{The median difference per epoch vs.\ time for the full catalog (black) and the A3 chip (red), which is known to have greater impact due to persistence. The higher amplitude of the A3 data is further indication that this phenomenon is due to persistence.}
    \label{fig:persistence}
\end{figure}

We consider two possible explanations for this phenomenon. The first is detector persistence, where lingering electrons in pixels receiving large signal can create lingering after-images from previous integrations. This has been previously studied on the NIRCam instrument, and the NIRCam A3 chip is known to experience the most severe persistence. \autoref{fig:persistence} again shows the median offsets for the whole catalog, as well as those for only the stars which fall on A3. The A3 subset shows a marginally more pronounced effect. Past studies of HST data have shown a similar phenomenon in WFC3/IR photometry, which is rightfully attributed to persistence \citep{Bajaj2019}. However, it is important to highlight that the timescales on which persistence is observed in HST are much longer than what we observe here, despite the effect being of similar size. The JWST documentation\footnote{https://jwst-docs.stsci.edu/jwst-near-infrared-camera/nircam-performance/nircam-persistence} also reports a NIRCam persistence decay timescale on the order of hours.

Because of the difference in relevant timescales, it is more likely that this phenomenon is caused by observational burn-in, where some of the initial charge in the detector is captured by empty traps, causing a brief reduction in signal before an equilibrium is reached \citep{Regan2012}. In studies conducted on the Roman detectors, for example, burn-in equilibrium is reached in $\sim500s$, about the 4th or 5th group in the data presented here \citep{Mosby2020}. Each time the instrument dithers, the traps associated with the new source location pixels need to re-fill again, creating the ``reset''.

\begin{figure}[h]
    \centering
    \includegraphics[width=0.45\linewidth]{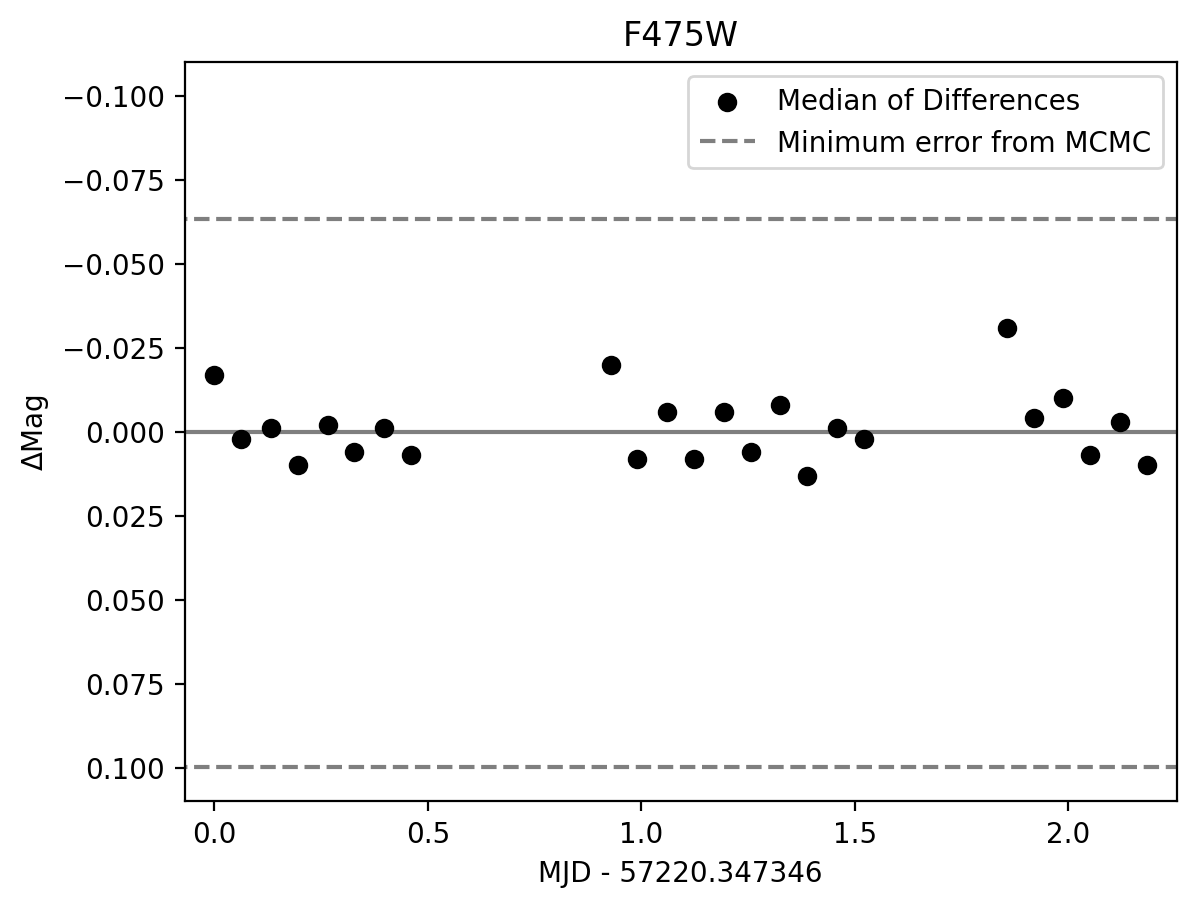}
    \includegraphics[width=0.45\linewidth]{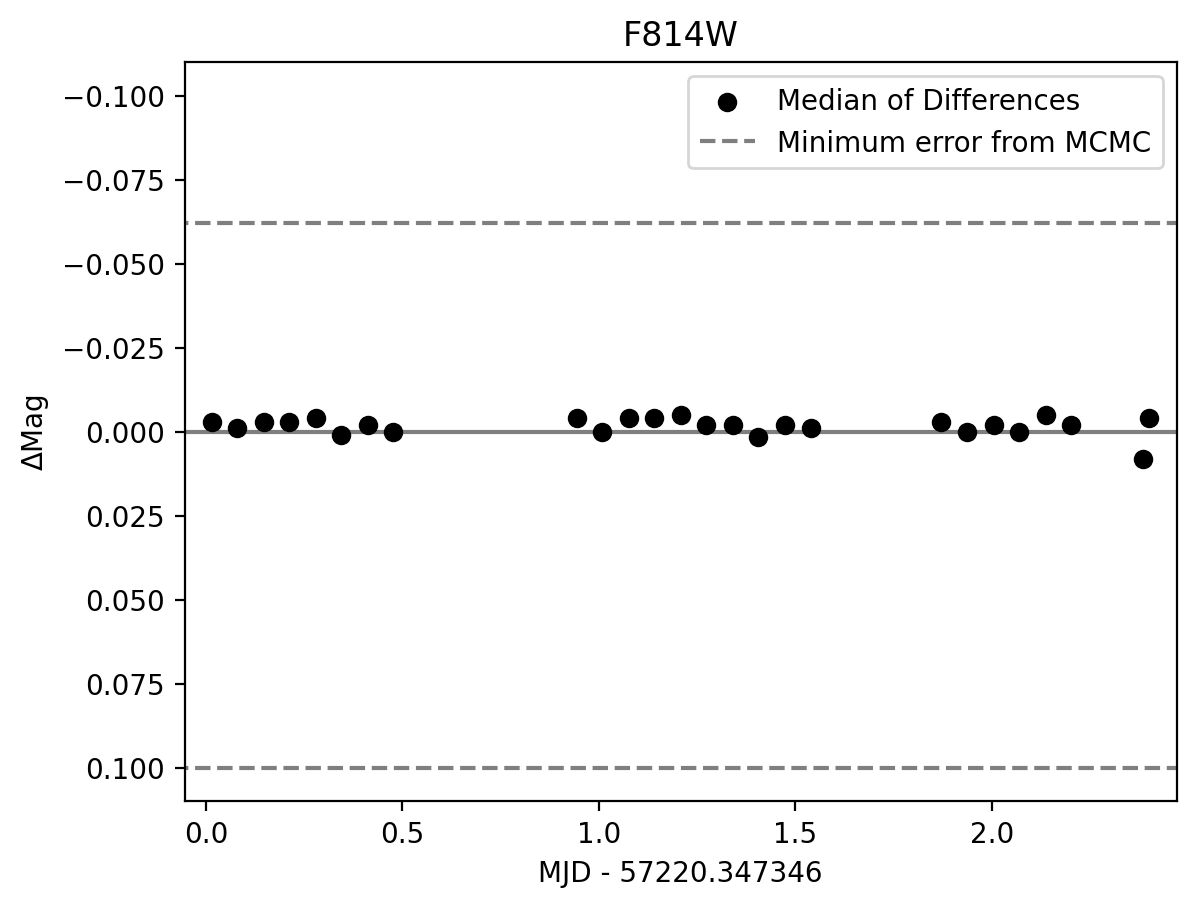}
    \caption{The median difference per epoch vs.\ time for both HST filter data sets (F475W, left; F814W, right). In both, the zero-point is indicated by a solid line, and the minimum (asymmetric) errors on the amplitude, as output by the template-fitting MCMC, are indicated by dashed lines.}
    \label{fig:HSTzp}
\end{figure}

When conducting the same analysis on the Hubble data, we find the scatter to be comparable or greater. This is unsurprising, given that the HST operates in a more challenging environment than JWST, and therefore more difficult to keep stable, despite being physically smaller. \autoref{fig:HSTzp} shows these results. As a charge-coupled device, the ACS does not experience the burn-in/persistence effect. We suspect that the scatter in the ACS observations are a result of so-called telescope ``breathing": a known change in focus position due to temperature fluctuations, which impacts PSF fitting.

\newpage
\section{Extended Data Tables}\label{sec:tables}

\begin{deluxetable}{cccccccccccc}[h]
\caption{HST Epoch Photometry\label{tab:HSTepoch}}

\tablehead{\colhead{MJD} & \colhead{Filter} & \colhead{ }& \colhead{24909} & \colhead{24909} & \colhead{25157} & \colhead{25157} & \colhead{25516} & \colhead{25516} & \colhead{25844} & \colhead{25844}\\
\colhead{} & \colhead{Number} & \colhead{ }& \colhead{Mag.} & \colhead{Unc.} & \colhead{Mag.} & \colhead{Unc.} & \colhead{Mag.} & \colhead{Unc.} & \colhead{Mag.} & \colhead{Unc.}\\
\colhead{$\mathrm{d}$} & \colhead{ } & \colhead{ }& \colhead{$\mathrm{mag}$} & \colhead{$\mathrm{mag}$} & \colhead{$\mathrm{mag}$} & \colhead{$\mathrm{mag}$} & \colhead{$\mathrm{mag}$} & \colhead{$\mathrm{mag}$} & \colhead{$\mathrm{mag}$} & \colhead{$\mathrm{mag}$}}

\startdata
57220.347346 & 0 & ... & 25.955 & 0.05 & 26.209 & 0.057 & 25.861 & 0.044 & 25.806 & 0.046 & ... \\
57220.410298 & 0 & ... & 26.035 & 0.053 & 26.268 & 0.059 & 26.027 & 0.049 & 25.74 & 0.046 & ... \\
57220.479904 & 0 & ... & 25.982 & 0.051 & 26.175 & 0.055 & 26.092 & 0.052 & 26.038 & 0.051 & ... \\
57220.542855 & 0 & ... & 25.82 & 0.047 & 25.902 & 0.049 & 26.131 & 0.053 & 25.996 & 0.051 & ... \\
57220.612462 & 0 & ... & 25.758 & 0.044 & 25.47 & 0.035 & 26.092 & 0.051 & 26.125 & 0.055 & ... \\
...& ... & ... & ... & ... & ... & ... & ... & ... & ... & ... & ... \\
57222.415587 & 1 & ... & 25.007 & 0.042 & 25.064 & 0.042 & 25.114 & 0.041 & 25.182 & 0.045 & ... \\
57222.484349 & 1 & ... & 25.001 & 0.043 & 25.224 & 0.055 & 25.053 & 0.043 & 25.179 & 0.048 & ... \\
57222.548156 & 1 & ... & 25.171 & 0.053 & 25.116 & 0.044 & 25.144 & 0.043 & 25.221 & 0.047 & ... \\
57222.731987 & 1 & ... & 25.338 & 0.056 & 25.275 & 0.055 & 25.092 & 0.046 & 24.893 & 0.042 & ... \\
57222.7473 & 1 & ... & 25.15 & 0.047 & 25.187 & 0.049 & 25.097 & 0.045 & 24.889 & 0.039 & ...\\
\enddata

\tablecomments{An abbreviated version of the full HST epoch photometry data, for the final 101 RR Lyrae. In the ``Filter Number'' column, 0 and 1 correspond to observations taken with the F475W and F814W filters, respectively. Data are also available in full in machine-readable format.}
\end{deluxetable}

 \begin{deluxetable}{cccccccccccc}[h]
\caption{JWST Epoch Photometry\label{tab:JWSTepoch}}

\tablehead{\colhead{MJD} & \colhead{Filter} & \colhead{} & \colhead{1330172} & \colhead{1330172} & \colhead{1330186} & \colhead{1330186} & \colhead{1330202} & \colhead{1330202} & \colhead{1330208} & \colhead{1330208}& \colhead{ }\\ 
\colhead{} & \colhead{Number} & \colhead{} & \colhead{Mag.} & \colhead{Unc.} & \colhead{Mag.} & \colhead{Unc.} & \colhead{Mag.} & \colhead{Unc.} & \colhead{Mag.} & \colhead{Unc.}& \colhead{ }\\ 
\colhead{$\mathrm{d}$} & \colhead{ } & \colhead{ } & \colhead{$\mathrm{mag}$} & \colhead{$\mathrm{mag}$} & \colhead{$\mathrm{mag}$} & \colhead{$\mathrm{mag}$} & \colhead{$\mathrm{mag}$} & \colhead{$\mathrm{mag}$} & \colhead{$\mathrm{mag}$} & \colhead{$\mathrm{mag}$}& \colhead{ }}
\startdata
59783.977914 & 0 & ... & 25.254 & 0.031 & 25.237 & 0.031 & 24.924 & 0.026 & 24.876 & 0.025 & ... \\
59783.987731 & 0 & ... & 25.245 & 0.031 & 25.175 & 0.03 & 24.835 & 0.025 & 24.861 & 0.025 & ... \\
59783.997548 & 0 & ... & 25.141 & 0.029 & 25.029 & 0.027 & 24.822 & 0.025 & 24.899 & 0.025 & ... \\
59784.007366 & 0 & ... & 25.089 & 0.028 & 24.878 & 0.025 & 24.751 & 0.025 & 24.9 & 0.025 & ... \\
59784.017183 & 0 & ... & 24.999 & 0.026 & 24.761 & 0.025 & 24.765 & 0.025 & 24.89 & 0.025 & ... \\
...& ... & ... & ... & ... & ... & ... & ... & ... & ... & ... & ... \\
59784.567193 & 1 & ... & 24.626 & 0.028 & 24.655 & 0.028 & 24.529 & 0.026 & 24.365 & 0.025 & ... \\
59784.57701 & 1 & ... & 24.615 & 0.027 & 24.5 & 0.026 & 24.501 & 0.026 & 24.357 & 0.025 & ... \\
59784.586827 & 1 & ... & 24.61 & 0.027 & 24.451 & 0.025 & 24.429 & 0.025 & 24.336 & 0.025 & ... \\
59784.596645 & 1 & ... & 24.654 & 0.028 & 24.405 & 0.025 & 24.408 & 0.025 & 24.354 & 0.025 & ... \\
59784.606462 & 1 & ... & 24.587 & 0.027 & 24.388 & 0.025 & 24.426 & 0.025 & 24.321 & 0.025 & ...\\
\enddata
\tablecomments{An abbreviated version of the full JWST RR Lyrae candidate epoch photometry data. In the ``Filter Number'' column, 0 and 1 correspond to observations taken with the F090W and F150W filters, respectively. Data are also available in full in machine-readable format.}
\end{deluxetable}

\begin{splitdeluxetable*}{ccccccccBccccccBcccccc}
\caption{HST Characteristic Parameters and MCMC Template-Fitting Output\label{tab:HSTfinalfits}}

\tablehead{\colhead{HST ID} & \colhead{RA} & \colhead{Dec} & \colhead{$\chi^2$} & \colhead{Type} & \colhead{Period} & \colhead{eP-} & \colhead{eP+} & \colhead{F475W Amp} & \colhead{F475W eAmp-} & \colhead{F475W eAmp+} & \colhead{F475W Mag} & \colhead{F475W eMag-} & \colhead{F475W eMag+} & \colhead{F814W Amp} & \colhead{F814W eAmp-} & \colhead{F814W eAmp+} & \colhead{F814W Mag} & \colhead{F814W eMag-} & \colhead{F814W eMag+}\\ 
\colhead{ } & \colhead{$\mathrm{deg}$} & \colhead{$\mathrm{deg}$} & \colhead{ } & \colhead{ } & \colhead{$\mathrm{d}$} & \colhead{$\mathrm{d}$} & \colhead{$\mathrm{d}$} & \colhead{$\mathrm{mag}$} & \colhead{$\mathrm{mag}$} & \colhead{$\mathrm{mag}$} & \colhead{$\mathrm{mag}$} & \colhead{$\mathrm{mag}$} & \colhead{$\mathrm{mag}$} & \colhead{$\mathrm{mag}$} & \colhead{$\mathrm{mag}$} & \colhead{$\mathrm{mag}$} & \colhead{$\mathrm{mag}$} & \colhead{$\mathrm{mag}$} & \colhead{$\mathrm{mag}$}}

\startdata
... & ... & ... & ... & ... & ... & ... & ... & ... & ... & ... & ... & ... & ... & ... & ... & ... & ... & ... & ... \\
24688 & 0.48014 & -15.50896 & 2.347 & ab & 0.579 & 0.009 & 0.01 & 1.014 & 0.144 & 0.14 & 25.783 & 0.047 & 0.046 & 0.475 & 0.152 & 0.154 & 25.033 & 0.041 & 0.042 \\
24772 & 0.487329 & -15.51667 & 2.952 & ab & 0.65 & 0.026 & 0.032 & 0.537 & 0.155 & 0.148 & 25.868 & 0.047 & 0.048 & 0.265 & 0.128 & 0.121 & 24.97 & 0.039 & 0.04 \\
24848 & 0.495201 & -15.5068 & 2.026 & ab & 0.595 & 0.018 & 0.022 & 0.595 & 0.169 & 0.158 & 25.876 & 0.05 & 0.052 & 0.225 & 0.139 & 0.136 & 25.002 & 0.042 & 0.04 \\
24884 & 0.475586 & -15.50041 & 2.083 & ab & 0.578 & 0.007 & 0.006 & 1.006 & 0.17 & 0.169 & 25.82 & 0.047 & 0.045 & 0.642 & 0.158 & 0.153 & 25.028 & 0.039 & 0.041 \\
24909 & 0.486729 & -15.50837 & 8.158 & c & 0.375 & 0.009 & 0.01 & 0.471 & 0.131 & 0.133 & 25.728 & 0.042 & 0.043 & 0.26 & 0.143 & 0.134 & 25.092 & 0.044 & 0.04 \\
25157 & 0.490591 & -15.49442 & 4.117 & ab & 0.564 & 0.009 & 0.009 & 0.741 & 0.152 & 0.146 & 25.886 & 0.048 & 0.046 & 0.286 & 0.145 & 0.13 & 25.074 & 0.042 & 0.042 \\
25516 & 0.5044 & -15.51125 & 1.729 & ab & 0.668 & 0.033 & 0.041 & 0.402 & 0.128 & 0.159 & 25.941 & 0.05 & 0.044 & 0.297 & 0.072 & 0.114 & 25.024 & 0.039 & 0.038 \\
25844 & 0.475218 & -15.49818 & 1.278 & ab & 0.618 & 0.018 & 0.02 & 0.66 & 0.206 & 0.2 & 25.878 & 0.051 & 0.048 & 0.407 & 0.156 & 0.141 & 25.041 & 0.042 & 0.042 \\
27165 & 0.493828 & -15.50734 & 1.879 & ab & 0.574 & 0.009 & 0.01 & 1.007 & 0.163 & 0.176 & 26.003 & 0.056 & 0.055 & 0.545 & 0.153 & 0.153 & 25.086 & 0.038 & 0.04 \\
28037 & 0.503707 & -15.50778 & 1.607 & ab & 0.583 & 0.01 & 0.008 & 0.783 & 0.172 & 0.18 & 26.029 & 0.054 & 0.052 & 0.359 & 0.179 & 0.167 & 25.14 & 0.044 & 0.045\\
\enddata

\tablecomments{An abbreviated version of the full HST template-fitting results for the final 101 RR Lyrae. Data are also available in full in machine-readable format.}
\end{splitdeluxetable*}

\begin{splitdeluxetable*}{cccccccccBccccccBcccccc}
\caption{JWST Characteristic Parameters and MCMC Template-Fitting Output\label{tab:JWSTfinalfits}}

\tablehead{\colhead{JWST ID} & \colhead{HST ID} & \colhead{RA} & \colhead{Dec} & \colhead{$\chi^2$} & \colhead{Type} & \colhead{Period} & \colhead{eP-} & \colhead{eP+} & \colhead{F090W Amp} & \colhead{F090W eAmp-} & \colhead{F090W eAmp+} & \colhead{F090W Mag} & \colhead{F090W eMag-} & \colhead{F090W eMag+} & \colhead{F150W Amp} & \colhead{F150W eAmp-} & \colhead{F150W eAmp+} & \colhead{F150W Mag} & \colhead{F150W eMag-} & \colhead{F150W eMag+}\\ \colhead{ } & \colhead{ } & \colhead{$\mathrm{deg}$} & \colhead{$\mathrm{deg}$} & \colhead{ } & \colhead{ } & \colhead{$\mathrm{d}$} & \colhead{$\mathrm{d}$} & \colhead{$\mathrm{d}$} & \colhead{$\mathrm{mag}$} & \colhead{$\mathrm{mag}$} & \colhead{$\mathrm{mag}$} & \colhead{$\mathrm{mag}$} & \colhead{$\mathrm{mag}$} & \colhead{$\mathrm{mag}$} & \colhead{$\mathrm{mag}$} & \colhead{$\mathrm{mag}$} & \colhead{$\mathrm{mag}$} & \colhead{$\mathrm{mag}$} & \colhead{$\mathrm{mag}$} & \colhead{$\mathrm{mag}$}}

\startdata
... & ... & ... & ... & ... & ... & ... & ... & ... & ... & ... & ... & ... & ... & ... & ... & ... & ... & ... & ... & ... \\
1330172 & 25844 & 0.475188 & -15.49816 & 1.358 & ab & 0.618 & 0.018 & 0.02 & 0.573 & 0.155 & 0.16 & 25.04 & 0.039 & 0.039 & 0.366 & 0.186 & 0.354 & 24.548 & 0.076 & 0.108 \\
1330202 & 24619 & 0.474795 & -15.49981 & 2.056 & ab & 0.587 & 0.008 & 0.008 & 0.451 & 0.121 & 0.128 & 24.951 & 0.079 & 0.05 & 0.191 & 0.123 & 0.128 & 24.501 & 0.049 & 0.055 \\
1330208 & 20325 & 0.480818 & -15.50552 & 1.11 & ab & 0.607 & 0.017 & 0.024 & 0.531 & 0.146 & 0.179 & 25.023 & 0.076 & 0.064 & 0.29 & 0.075 & 0.089 & 24.435 & 0.03 & 0.034 \\
1330263 & 25157 & 0.49056 & -15.49441 & 1.716 & ab & 0.564 & 0.009 & 0.009 & 0.381 & 0.228 & 0.34 & 25.023 & 0.094 & 0.067 & 0.144 & 0.107 & 0.298 & 24.466 & 0.042 & 0.074 \\
1330483 & 24606 & 0.483601 & -15.5044 & 0.867 & c & 0.38 & 0.006 & 0.007 & 0.264 & 0.074 & 0.074 & 25.105 & 0.024 & 0.025 & 0.073 & 0.051 & 0.079 & 24.66 & 0.031 & 0.029 \\
1516921 & 22827 & 0.508518 & -15.4981 & 1.355 & ab & 0.631 & 0.007 & 0.008 & 0.343 & 0.125 & 0.134 & 24.96 & 0.04 & 0.036 & 0.127 & 0.094 & 0.23 & 24.342 & 0.046 & 0.07 \\
1516944 & 20693 & 0.493523 & -15.50062 & 0.723 & ab & 0.631 & 0.009 & 0.011 & 0.362 & 0.127 & 0.132 & 24.901 & 0.066 & 0.059 & 0.215 & 0.083 & 0.081 & 24.387 & 0.034 & 0.034 \\
1517050 & 23440 & 0.507812 & -15.50092 & 1.173 & ab & 0.613 & 0.007 & 0.008 & 0.451 & 0.132 & 0.139 & 24.958 & 0.068 & 0.062 & 0.264 & 0.081 & 0.08 & 24.444 & 0.035 & 0.035 \\
1517067 & 23808 & 0.493475 & -15.49482 & 1.262 & ab & 0.613 & 0.009 & 0.011 & 0.189 & 0.099 & 0.096 & 24.996 & 0.037 & 0.03 & 0.235 & 0.135 & 0.201 & 24.527 & 0.058 & 0.072 \\
1517099 & 23715 & 0.496747 & -15.49512 & 1.84 & ab & 0.591 & 0.008 & 0.008 & 0.137 & 0.091 & 0.107 & 24.992 & 0.037 & 0.032 & 0.147 & 0.094 & 0.114 & 24.527 & 0.053 & 0.049\\
... & ... & ... & ... & ... & ... & ... & ... & ... & ... & ... & ... & ... & ... & ... & ... & ... & ... & ... & ... & ...\\
\enddata

\tablecomments{An abbreviated version of the JWST overlap-region RR Lyrae MCMC template-fitting output. The periods and associated uncertainties reported here are those obtained from the HST template-fitting. Data are also available in full in machine-readable format.}
\end{splitdeluxetable*}

\bibliography{bibliography}{}
\bibliographystyle{aasjournalv7}

\end{document}